\documentclass[onecolumn]{IEEEtran}
\IEEEoverridecommandlockouts
\usepackage{cite}
\usepackage{amsmath,amssymb,amsfonts}
\usepackage{graphicx}
\usepackage{textcomp}
\usepackage{xcolor}
\usepackage{subfigure}
\usepackage{cases}
\usepackage{soul} 
\usepackage{multirow}
\usepackage{multicol}
\usepackage{rotating}
\usepackage{booktabs}
\usepackage{mathtools}
\usepackage{breqn}
\usepackage{algorithm} 
\usepackage{algpseudocode}
\usepackage[version=3]{mhchem} 
\usepackage{hyperref}
\def\BibTeX{{\rm B\kern-.05em{\sc i\kern-.025em b}\kern-.08em
    T\kern-.1667em\lower.7ex\hbox{E}\kern-.125emX}}

\algnewcommand{\Inputs}[1]{%
  \State \textbf{Inputs:}
  \Statex \hspace*{\algorithmicindent}\parbox[t]{.8\linewidth}{\raggedright #1}
}
\algnewcommand{\Initialize}[1]{%
  \State \textbf{Initialize:}
  \Statex \hspace*{\algorithmicindent}\parbox[t]{.8\linewidth}{\raggedright #1}
}

\title{Utilizing Neurons for Digital Logic Circuits: A Molecular Communications Analysis}

\author{Geoflly L. Adonias$^{1}$, Anastasia Yastrebova$^{2}$,  Michael T. Barros$^{1}$, Yevgeni Koucheryavy$^{2}$~\IEEEmembership{Senior Member, IEEE},\\Frances~Cleary$^{1}$, and Sasitharan~Balasubramaniam$^{1}$,~\IEEEmembership{Senior Member, IEEE}%

\thanks{$^{1}$G. L. Adonias, M. T. Barros, F. Cleary and Sasitharan Balasubramaniam are with the Telecommunications Software \& Systems Group, Waterford Institute of Technology, Waterford, Ireland, e-mails: \{gadonias, mbarros, fcleary, sasib\}@tssg.org.}%
\thanks{$^{2}$A. Yastrebova and Y. Koucheryavy are with the Faculty of Information Technology and Communication Sciences, Tampere University, Tampere, Finland, e-mails: anastasia.yastrebova@tuni.fi, yk@cs.tuni.fi.}%
\thanks{This work is supported by the Science Foundation Ireland (SFI) CONNECT Project under grant no. 1R/RC/2077.}%
\thanks{Submitted Manuscript.}}

\begin{document}
\maketitle

\begin{abstract}
With the advancement of synthetic biology, several new tools have been conceptualized over the years as alternative treatments for current medical procedures. Most of those applications are applied to various chronic diseases. This work investigates how synthetically engineered neurons can operate as digital logic gates that can be used towards bio-computing for the brain.  We quantify the accuracy of logic gates under high firing rates amid a network of neurons and by how much it can smooth out uncontrolled neuronal firings. To test the efficacy of our method, simulations composed of computational models of neurons connected in a structure that represents a logic gate are performed. The simulations demonstrated the accuracy of performing the correct logic operation, and how specific properties such as the firing rate can play an important role in the accuracy. As part of the analysis, the Mean squared error is used to quantify the quality of our proposed model and predicting the accurate operation of a gate based on different sampling frequencies. As an application, the logic gates were used to trap epileptic seizures in a neuronal network, where the results demonstrated the effectiveness of reducing the firing rate. Our proposed system has the potential for computing numerous neurological conditions of the brain.
\end{abstract}


\section{Introduction}
\label{sec:introduction}

It has been over a decade since \textit{Molecular Communications} was introduced as a new communication paradigm aiming to conceptualize and build communication systems inspired by natural biological processes~\cite{nakano2017, akyildiz2015internet, Felicetti2014, FELICETTI201627}. One of them is known as \textit{neuro-spike communication}~\cite{Balevi2013}, where information is transferred between two neurons through electro-chemical process which  triggers an electrical impulse called \textit{action potentials}, i.e. spikes. This information can not only be encoded~\cite{Thorpe2001, Rolls2011, Rinkus2010, Rao2011, Luczak2015, Zeldenrust2018} but also modulated~\cite{Adonias2018, Billimoria5910, Yi2014} mimicking a traditional communication system and potentially presenting itself as a tool for cognitive enhancement and treatment of neurodegenerative diseases~\cite{Choe2016, pavon2014}.

Synthetic Biology has enabled us to engineer cells that can modify or inherit new functions~\cite{Loscri2015, Loscri2018, Egan2019}. While the majority of focus of synthetic biology has been for prokaryotic cells, there is potential to engineer mammalian cells such as neurons and astrocytes that can be synthetically engineered to enable control of their dynamic behaviour and functionality, aiming at correction of abnormalities at a cellular level. With the advance of synthetic biology and nano-scale networks~\cite{akyildiz2008nanonetworks}, many components ranging from logic gates~\cite{Adonias2019, hasty2002engineered, Goldental2014, Vogels10786} to integrated circuits such as oscillators~\cite{morse2002time} have also emerged. To date, there has not been an application of synthetic logic gates for neurological diseases. 

Information processing in the brain involves the propagation of action potentials through countless numbers of specific neuronal networks. This enables the brain to process various types of information that can range from controlling the functions of organs within an organism to coding and storing long-term memory, as examples. A synchronous uncontrolled firing of spikes in large regions of the brain can occur spontaneously can can be related to  neurological diseases, and one example is epilepsy~\cite{jirsa2014}. Based on this, an example application of synthetic logic gates is to use them as digital filters and when positioned in between neurons as part of the network, they can play an important role in smoothing out uncontrolled neuronal firing and consequently reducing the effects of seizures. To the best of our knowledge, there are no works on neuronal logic gates that uses such  models and investigate their effects in scenarios affected by neurological disorders.

\begin{figure}[t]
    \centering
    \includegraphics[width=0.4\columnwidth]{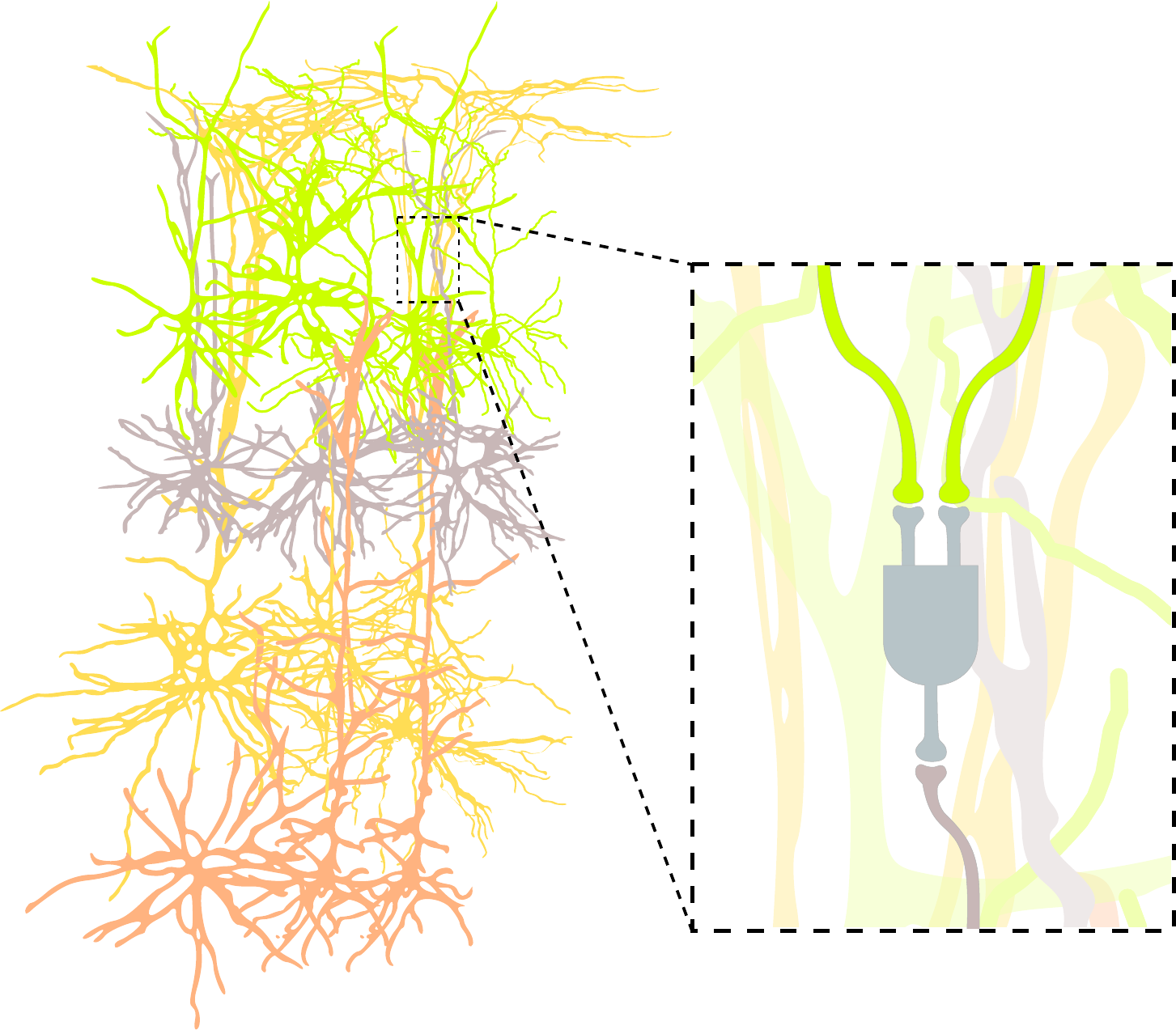}
    \caption{Neuronal logic gate inside a cortical column.}
    \label{fig:gatewithin}
\end{figure}

In this paper, we build on top of our previous work~\cite{Adonias2019} and analyze the effects of inserting neuronal logic gates in a network of neurons. The main contributions of this paper are as follows:

\begin{itemize}
    \item \textbf{Neuronal logic gates are built, controlled and simulated within large neuronal networks using computational models of neurons}~\cite{Markram2015}. We use three models of neuron cells to create a single synthetic logic gate capable of performing logic operations at a cellular level.
    \item \textbf{Analysis of performance for the gates simulated in isolation and inside a network of neurons.} We analyze the dynamic behaviour of neuronal communications that could affect the operation of the gate and consequently the network, quantified in terms of accuracy.
    \item \textbf{Proposal of a queuing model for the input and gating of action potentials as units of information.} The model is capable of predicting the accuracy of the synthetic gates, and this is validated using mean squared error (MSE) as a function of the inter-spike interval (ISI) at different sampling frequencies.
\end{itemize}

The remainder of this paper is organized as follows, in Section~\ref{sec:related_work}, a review of the current literature is provided. The construction of neuronal logic gates, their diverse types and the queue-theoretical analysis are discussed in Section~\ref{sec:neuronalgatescircuits}. Section~\ref{sec:simulation} contains all details regarding the simulation, network connectivity and its parameters and the results are presented. Finally, in Section~\ref{sec:conclusions} the conclusions for this work are presented.

\section{Related Work}
\label{sec:related_work}

In the 1940s, McCulloch and Pitts envisioned that the brain would be composed of units with logic gating capabilities~\cite{McCulloch1943}. During that time this did not have much impact in neuroscience simply because neuronal cells are much more dynamic than simple digital logic gates, but their proposal did positively impact advancements in artificial neural network and machine learning theories~\cite{Goldental2014}. The vision of creating logic operating engineered systems to interact between natural cells and engineered cells (i.e. bio-nanomachines) has the potential to create better alternatives for the treatment of diseases at the cellular level~\cite{Miyamoto2013}.

In light of numerous applications that can have an impact on biological systems~\cite{Wang2012}, researchers have been investigating how logic gates can not only help treat neurodegenerative diseases but also enhance biological computations performed by the brain. One example is the work of Vogels and Abbott~\cite{Vogels10786}, where they investigated the signal propagation in networks of integrate-and-fire models of neurons and found that by either strengthening or weakening specific synapses, different types of logic gates may arise within the network. This ``fine-tunning'' approach to synaptic connections may pave the way to more sophisticated systems capable of, for example, weakening connections between neurons involved in a seizure or even strengthening connections responsible for memory storage.

Goldental and colleagues~\cite{Goldental2014}, on the other hand, decided to build on top of the fact that the way neurons work and communicate with each other exhibit far richer dynamics and so they proposed the concept of dynamic logic gates that work as a function of their historical activities, interconnection profiles, as well as the frequency of stimulation at the input terminals.

Song \textit{et al}~\cite{SONG2016380} took a different approach, where they considered a ``third-party'' for the analysis and proposed that the interaction of astrocytes in a tripartite synapse may be able to control the logic gate performance of neurons. Although they are also working with non-neuronal cells, the neurons used are all of the same morphological and electrical types.

\section{Neuron Communication Background}

Neuronal network communications allow the propagation of spikes through a population of neurons transferring information inside the brain. Bio-computing approaches based on the communication of neurons will rely on these propagation behaviour and its relation to the neuron properties as well as the characterization of neuronal communications. Therefore, before presenting the model of logic gates using neurons (Section~\ref{sec:neuronalgatescircuits}), the morpho-electrical characteristics, the columnar and laminar organization properties of neurons as well as the compartmentalized Hodgkin-Huxley model for neuronal communications will be introduced. 

\subsection{Neuron Properties}
\label{subsec:neuron_properties}
Neuronal cells can be classified in terms of their morphology, electrophysiology, projections, position in the brain and the proteins and genes they express. The models of neurons used in this work, collected from~\cite{Markram2015}, are classified only based on their morphological and electrical properties (morpho-electrical characteristics) as well as which cortical layer they are from (columnar and laminar organization). The classification method used in this work is detailed below.

\subsubsection{Morpho-electrical Characteristics}

Well-established features in the soma of the cell and its dendritic and axonal arbours are sufficient for the classification of different morphological cells. In terms of size, cortical neurons can be categorized as small neurons ($8-16\,\mu m$) along with neurons from the hippocampus, olfactory bulb and dorsal horn. Axonal features play a major role in distinguishing inhibitory types while excitatory types can be better identified by their dendritic features. On the other hand, the shape of the soma of cortical cells can help further categorize them into three major types~\cite{wirdatmadja2019analysis, Markram2015, Eckenstein2286}: 

\begin{enumerate}
    \item \textit{Pyramidal cells} which are predominantly found in layers III and V and have a distinctive pyramidal shape. Its size (height $\times$ width) varies between $12-100\,\mu m \times 10-60\,\mu m$,

    \item \textit{Granule cells} can be found mostly in layers II and IV. The ones in IV receive inputs from the thalamus and pass information forward to cells in all other layers except to layer I. Its size range (height $\times$ width) is $15-30\,\mu m \times 10-15\,\mu m$.

    \item  \textit{Fusiform cells} mainly populates in layer VI and have a flattened spindled-shaped form. They send information from the cortex back to the thalamus and some of its dendrites elongate up to layer I. The size of the soma for this type of cell varies between $15-30\,\mu m \times 10-15\,\mu m$~\cite{wirdatmadja2019analysis}. 
\end{enumerate}

Different morphological types (m-types) of cells can have diverse firing patterns. These patterns are generated in response to the injection of step currents in cortical neurons. From the 11 different electrical types (e-types) identified by Markram et al~\cite{Markram2015}, all m-types used in this work are burst Non-accommodating (bNAC) e-types.

\subsubsection{Columnar and Laminar Organization of the Cortex}
\label{subsec:cortical_layers}

The cerebral cortex is composed of neurons arranged into six horizontally and dispersed layers. These layers have different characteristics such as thickness, size, cell type and cell density showing a ``laminar'' organization and subdividing the cortex into disparate regions and areas. These layers are known as (1) Molecular layer, (2) External granular layer, (3) Pyramidal layer, (4) Inner granular layer, (5) Ganglionic layer and (6) Multiform layer.

Despite the horizontal layering, cortical regions display vertical connections that are of prime importance and take two forms: mini-columns (also called, micro-columns) with approximately $30-50$ $\mu$m in diameter and when activated by peripheral stimuli, it generates the macro-columns, with a diameter of approximately $0.4-0.5$~mm~\cite{Peters2010}.

\subsection{Neuronal Communications}
\label{subsec:Neuronal_comm}
\subsubsection{Neuron-to-neuron Communication}
\label{subsubsec:nncomm}

Communication between neurons is performed through electro-chemical synapses. Action potentials travel down the axon of the pre-synaptic cell and by the time it reaches the axon terminal, it stimulates the release of synaptic vesicles inside the synaptic cleft. These vesicles contain neurotransmitters that bind to neuro-receptors in the dendrites of the post-synaptic cell, on the other end of the synaptic cleft, either depolarizing the membrane. The depolarization starts in a potential state of approximately $-65$ mV and moves up to the point it reaches a threshold which is high enough to trigger the initiation of an action potential (excitatory) or polarizing the membrane even more, which in turn blocks the postsynaptic cell of firing any spikes (inhibitory)~\cite{Adonias2019, Mishra2019}. In larger networks, the balance between inhibitory and excitatory connections helps in encoding information through the neuronal network~\cite{Zhou2018}.

After the membrane potential reaches its maximum peak of depolarization, it starts to repolarize itself towards its resting potential  right after a spike is fired. The potential gets hyperpolarized for a very short period which is known as \textit{refractoriness} and can be subdivided into \textit{absolute} and \textit{relative}. During the absolute refractory period (ARP), the cell is unable to fire again regardless of how strong the stimuli are and it takes about $1-2$ ms followed by the relative refractory period (RRP) during which a cell can fire again if the applied stimulus is stronger than it was when applied at its resting state~\cite{Mishra2019}.

\begin{figure}[ht]
    \centering
    \includegraphics[width=0.5\columnwidth]{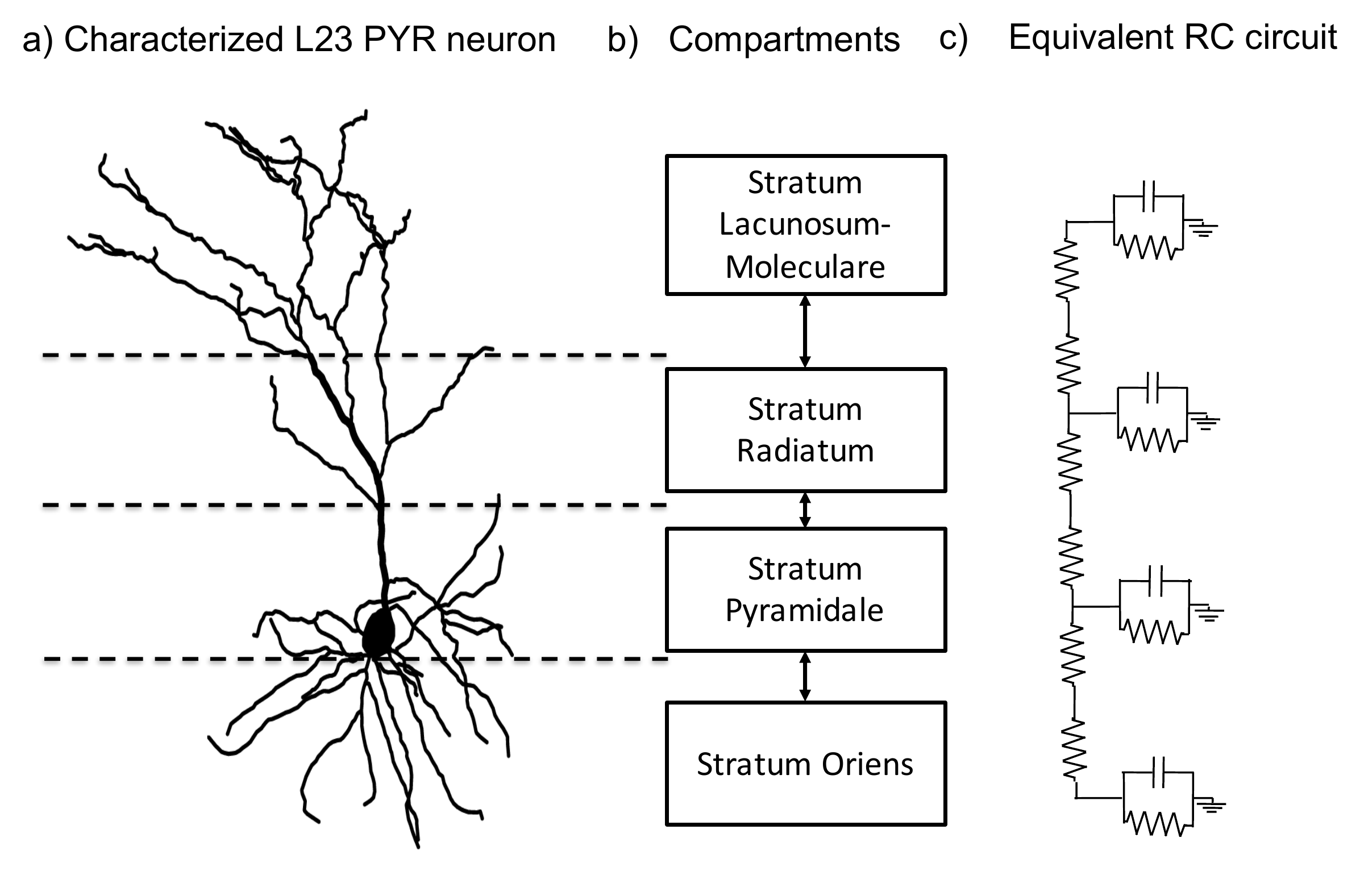}
    \caption{Cell comparmentalization; a) A morphological structure of a layer 2/3 pyramidal neuron, b) Compartment model of a Layer 2 pyramidal neuron including 4 compartments: stratum oriens, stratum pyramidale, stratum radiatum and stratum lacunosum-moleculare, c) The equivalent RC circuit module.}
    \label{fig:compartmenthh}
\end{figure}

To model such a complex system, we present a simplification of the model that is used in the NEURON simulator~\cite{Carnevale2009, Hines2009} based on the compartmentalized Hodgkin-Huxley model. In this model, the cell is broken down into $J$ equal length parts, and the spike propagation is modelled in one compartment travelling to all others (Fig.~\ref{fig:compartmenthh}).

We can describe a single compartment model with the following proposed by Pospischil \textit{et al}~\cite{pospischil2008minimal}

\begin{equation}
\label{eq:single_compartment}
    C_m \frac{\mbox{d}V}{\mbox{d}t} = -g_{leak}(V - E_{leak}) - I_{Na} - I_{K_d} - I_M - I_T - I_L,
\end{equation}

\noindent where $V$ is the membrane potential, $C_{m}$ is the specific capacitance of the membrane, $g_{leak}$ is the resting (leak) membrane conductance, $E_{leak}$ is its reversal potential. $I_{Na}$ and $I_{Kd}$ are the sodium and potassium currents responsible for action potentials respectively, $I_{M}$ is a slow voltage-dependent potassium current responsible for spike frequency adaptation, $I_{L}$ is a high-threshold calcium current and $I_{T}$ is a low-threshold calcium current. These voltage-dependent currents are variants of the same generic equation which is described as

\begin{equation}
\label{eq:vdcurrent}
    I_x = g_x m^M h^N (V - E_x),
\end{equation}

\noindent where the current $I_x$ is expressed as the product of the maximal conductance, $g_x$, activation ($m$) and inactivation ($h$) variables, respectively, and the difference between membrane potential $V$ and the reversal potential $E_{j}$. The gating of the channel is derived from the following first-order kinetic scheme

\begin{align*}
\ce{C <=>[\ce{\alpha (V)}][\ce{\beta (V)}] O\tag{\refstepcounter{equation}\theequation}},
\end{align*}

\noindent where $O$ and $C$ are the open and closed states of the gate, and $\alpha (V)$ and $\beta (V)$ are the transfer rates for each respective direction. The variables $m$ and $h$ represent the fraction of independent gates in the open state, following the conventional approach introduced by \cite{hodgkin1952quantitative} and stated as 

\begin{equation}
    \frac{\mbox{d}m}{\mbox{d}t} = \alpha_m (V) (1-m) - \beta_m (V) m,
\end{equation}

\begin{equation}
    \frac{\mbox{d}h}{\mbox{d}t} = \alpha_h (V) (1-h) - \beta_h (V) h.
\end{equation}

To consider conductance-based inputs to the neuron in (\ref{eq:single_compartment}), it is necessary to the effects from the propagation and reception of neurotransmitters from another neuron in the synaptic cleft. We present a simple model for the sake of brevity, in which the neurotransmitter-activated ion channels ($I_{\text{syn}}$) is represented as an explicitly time-dependent conductance ($g_{\text{syn}}$), and it is defined as

\begin{equation}
\label{eq:syn}
    I_{\text{syn}}=g_{\text{syn}}\,(V-E_{\text{syn}})\,,
\end{equation}

\noindent where the parameter $E_{\text{syn}}$ as well as $g_{\text{syn}}$ are used to describe the many different synapses types. Typically, $E_{\text{syn}} = -75$mV describes inhibitory synapses whereas $E_{\text{syn}} \approx 0$ describes excitatory ones. Based on this, $g_{\text{syn}}$ can be defined as through a superposition of exponentials

\begin{equation}
\label{eq:gsyn}
    g_{\text{syn}}=\sum_{f}\bar{g}_{\text{syn}}\,{\text{e}}^{-(t-t^{(f)})/\tau}%
\,H(t-t^{(f)})\,,
\end{equation}

\noindent where $\tau$ is a time constant, $t^{(f)}$ is the arrival time of a presynaptic action potential and $H(\cdot)$ is the Heaviside step function. The $t^{(f)}$ has a non-null value only when the membrane potential of the presynaptic compartment $V_{pre}$ crosses a threshold $th_{pre}$, indicating a spike has occurred. This threshold-crossing mechanism for spike propagation is known as \textit{event-based synapse} and it can be defined as

\begin{equation}
    \label{eq:syn_th}
        t^{(f)} = 
    \begin{dcases}
        t^{(f)}, & \text{if } V_{pre} \geq th_{pre}\\
        \emptyset, & \text{otherwise.}
    \end{dcases}
\end{equation}

\noindent 
This can be thought of each synaptic event as a number of neurotransmitters released and bound to the postsynaptic terminal~\cite{Akyildiz2019}.

Integrating (\ref{eq:single_compartment}) for compartments that synapses may occur requires a simple addition of (\ref{eq:syn}) on the left-hand side, as depicted below

\begin{equation}\label{eq:single_compartment_syn}
    C_m \frac{\mbox{d}V}{\mbox{d}t} = -g_{leak}(V - E_{leak}) - I_{Na} - I_{K_d} - I_{M} - I_{T} - I_{L} - I_{\text{syn}}(t).
\end{equation}

\begin{figure}[ht]
    \centering
    \includegraphics[width=.5\columnwidth]{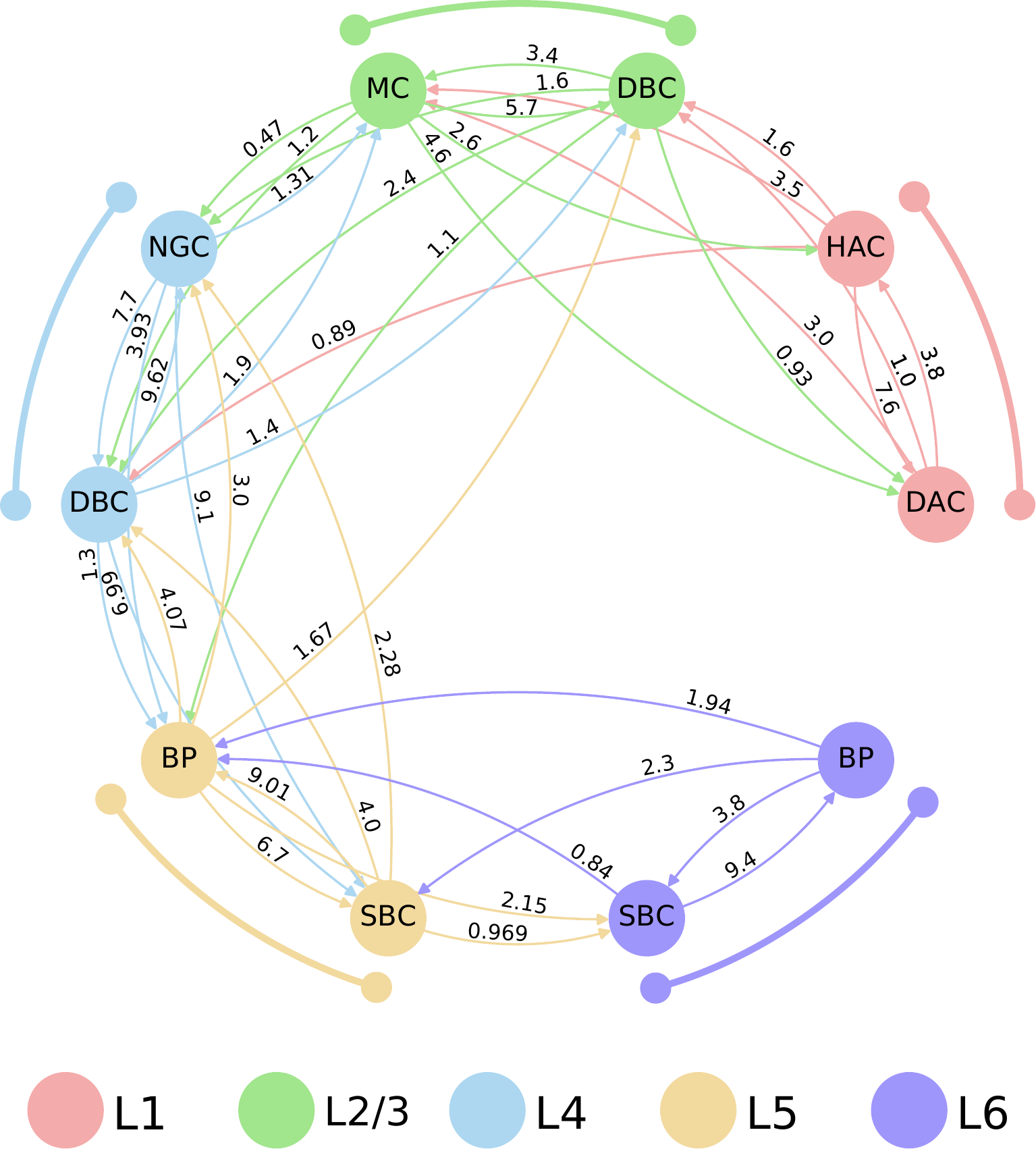}
    \caption{Graphical visualization of network connectivity and each connection probability (in percentage) between pairs of neurons.}
    \label{fig:connectivity}
\end{figure}

In Subsection~\ref{subsec:neuron_properties}, we presented the differences of morpho-electrical characteristics of neurons and their columnar and laminar organization that create a variety of neuronal networks with different types of cells. To incorporate these properties into the model presented in this section we need to use three different approaches. Their morphological properties will dictate the number of compartments of a cell type, which probably indicates that pyramidal, granule and fusiform cells will have different $V$ propagation patterns based on their different number of compartments. For example, in Fig.~\ref{fig:compartmenthh}, we divided a layer 2 pyramidal neuron into 4 compartments (stratum oriens, stratum pyramidale, stratum radiatum and stratum lacunosum-moleculare). By using the NEURON simulator, we can capture  morphological properties with precision through the validated models of Markram et al~\cite{Markram2015} in which compartments are already provided. In a similar manner, also using the models from~\cite{Markram2015}, it is necessary to change the following parameters in order to shape the electrical characteristics and obtain a type-specific spiking activity, other than the types already available: $m$, $h$, $\alpha(V)$, $\beta (V)$, $g_x$, $g_{syn}$, and initial values of $E_{leak}$ and $E_{syn}$. Lastly, based on the cell type, we define a network of excitatory neurons that consider the connection probabilities between them as defined in the \emph{Neocortical Microcircuit Collaboration Portal\footnote{\url{https://bbp.epfl.ch/nmc-portal/welcome}}}. We use a simple directed graph to capture this network connectivity pattern, as shown in Fig.~\ref{fig:connectivity}.

\subsubsection{Role of the Threshold in Event-based Synapses}

As aforementioned in Section~\ref{subsubsec:nncomm}, for a spike to be fired, the membrane potential of the cell compartment to reach a threshold during its depolarization state, the threshold for spike initiation varies with stimuli, cell type and the history of cell's activity. It is not yet clear what characteristics can cause this variability which may affect the performance of the gate. According to Platkiewicz and Brette~\cite{platkiewicz2010threshold}, even though the concept of spike threshold may be different for \textit{in vivo}, \textit{in vitro} and computational experiments, the threshold in brain cells depends on several parameters such as stimulus, type of cells, synaptic conductances and properties of ionic channels.

From a synaptic point of view, with each action potential arrival at the presynaptic terminal at time $t^{(f)}$, a specific number of neurotransmitters may be released into the synaptic cleft and has a probability of binding to the neuroreceptors at the postsynaptic cell. This release process is proportional to the shape and energy of the incoming action potential. An event-based synapse mimics this chemical process and sends an event with a synaptic weight to the postsynaptic cell that may trigger an action potential and consequently propagate information through the network.

To the best of our knowledge, there are no works that utilizes realistic models of neurons, and especially the neuron models proposed by Markram \textit{et al}~\cite{Markram2015}, where the gates are constructed from heterogeneous neuronal arrangements and controlled by their respective threshold for event-based synapses.

\section{Neuronal Digital Logic Gates and Circuits}
\label{sec:neuronalgatescircuits}
In this section we describe the construction of neuronal logic gates and how queue theory can be applied to neuronal circuits to predict and assess how the stimuli in the pre-synaptic terminal is being processed by the post-synatpic cell.
\subsection{Single Logic Gates}
\label{subsec:single_logic_gate_combination}

Eight neuronal logic gates were built, including five different OR gates and three different AND gates. The truth table for both of these types of gates is depicted in Table~\ref{tab:truth_table}. For an AND gate, both inputs must be non-null in order to have a non-null output. On the other hand, an OR gate can send out non-null output not only when both inputs are active but also when either one of them are active while the other is not.

\begin{table}[ht]
\centering
    \caption{Truth table for both gate types.}
    \label{tab:truth_table}
    \begin{tabular}{c|c|c|c}
        \hline
        \hline
        \multicolumn{4}{c}{Truth Table}\\
        \hline
        I$_{1}$ & I$_{2}$ & O$_{\textsc{AND}}$ & O$_{\textsc{OR}}$\\
        \hline
         0 & 0 & 0 & 0 \\
         0 & 1 & 0 & 1 \\
         1 & 0 & 0 & 1 \\
         1 & 1 & 1 & 1 \\
        \hline
        \hline
    \end{tabular}
\end{table}

\begin{figure}[t]
	\centering
	\subfigure[\label{fig:set_of_gates}]{\includegraphics[width=0.66\columnwidth]{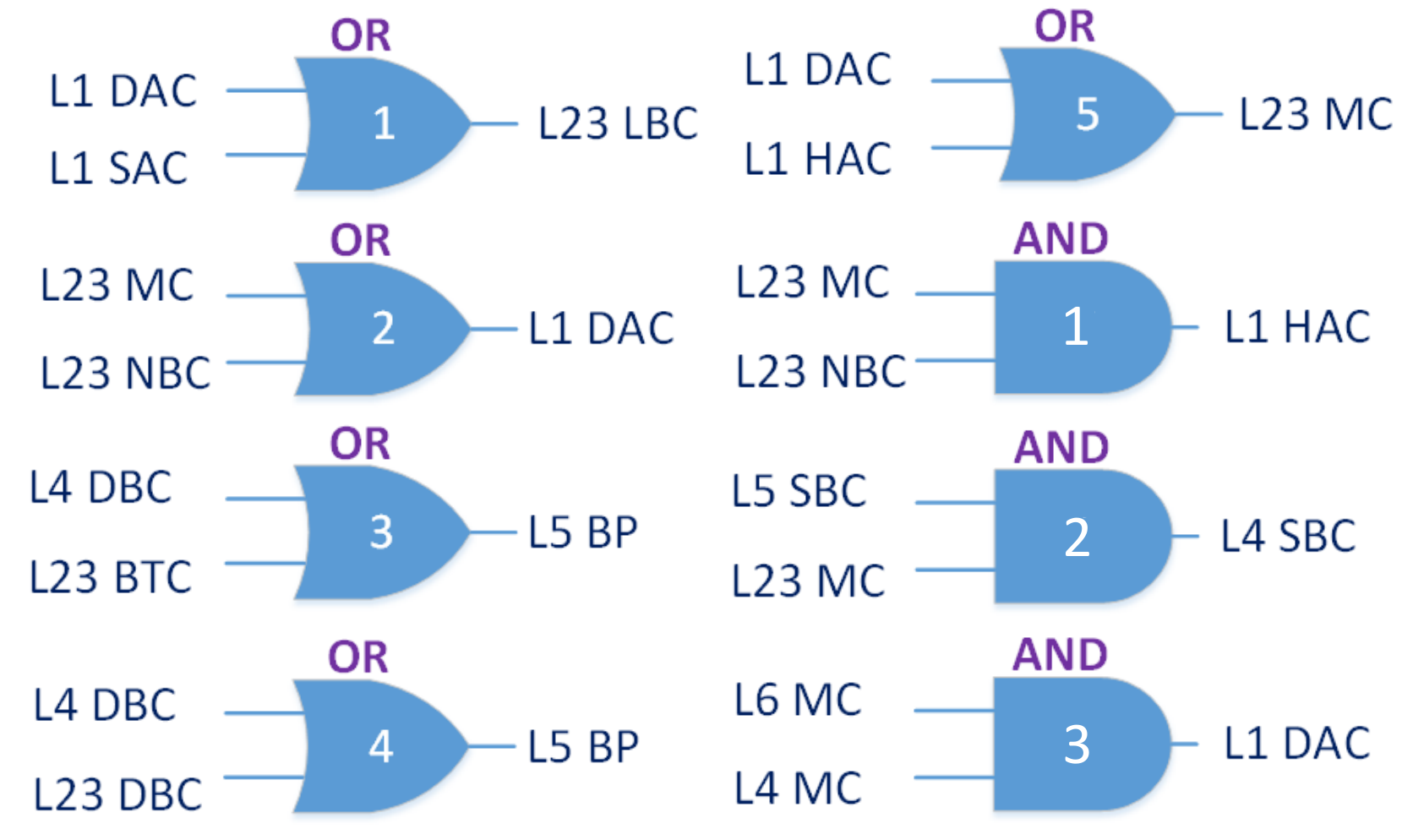}}
	~
   \subfigure[\label{fig:gate3d}]{\includegraphics[width=0.3\columnwidth]{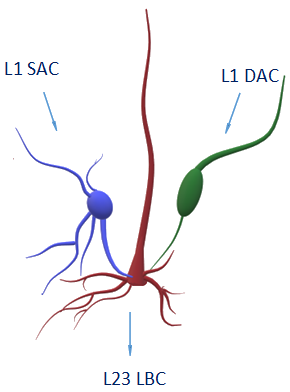}}
    \caption{Neuronal logic gates; (a) set of neuronal logic gates built using the models of neurons shown in Table~\ref{tab:cell_types} in a traditional representation and (b) potential real connection of neurons as a gate.\label{fig:single_logic}}
\end{figure}

All cell types used to build the gates are listed in Table~\ref{tab:cell_types}. For each gate, three different types of cells were arranged in a way that two of them should operate as the inputs of the gate and the third one as the output (Fig.~\ref{fig:gate3d}). The idea is to keep the inner connections of the gate, i.e. the connection between the inputs cells with the output cell, fixed at their default parameters and respective connection probabilities according to the type of cells being connected.

\begin{table}[ht]
\centering
    \caption{Types of cells used to build the gates.}
    \label{tab:cell_types}
    \begin{tabular}{c|c|c}
        \hline
        \hline
        \multicolumn{3}{c}{Cell Types}\\
        \hline
        \multirow{3}{*}{L1} & DAC & Descending Axon Cell \\
         & HAC & Horizontal Axon Cell \\ 
         & SAC & Small Axon Cell \\
        \hline
        \multirow{5}{*}{L2/3} & MC & Martinotti Cell \\
         & NBC & Nest Basket Cell \\
         & BTC & Bitufted Cell \\
         & DBC & Double Bouquet Cell \\
         & LBC & Large Basket Cell \\
        \hline
        \multirow{3}{*}{L4} & DBC & Double Bouquet Cell \\
         & SBC & Small Basket Cell \\
         & MC & Martinotti Cell \\
        \hline
        \multirow{2}{*}{L5} & BP & Bipolar Cell \\
         & SBC & Small Basket Cell \\
        \hline
        L6 & MC & Martinotti Cell \\
        \hline
        \hline
    \end{tabular}
\end{table}


Combinations of cells (as illustrated in Fig.~\ref{fig:set_of_gates}) were created largely based on their respective connection probabilities. Since the synaptic weight was kept at a fixed starting value, the higher the probability of two cells establishing a synapse, the higher the influence of the pre-synaptic cell on the post-synaptic cell. In this case, OR gates should have stronger inner connections  when compared to AND gates so we can achieve the desired behaviour, as presented in Table~\ref{tab:truth_table}.

\begin{figure}[t]
    \centering
    \includegraphics[width=0.5\columnwidth]{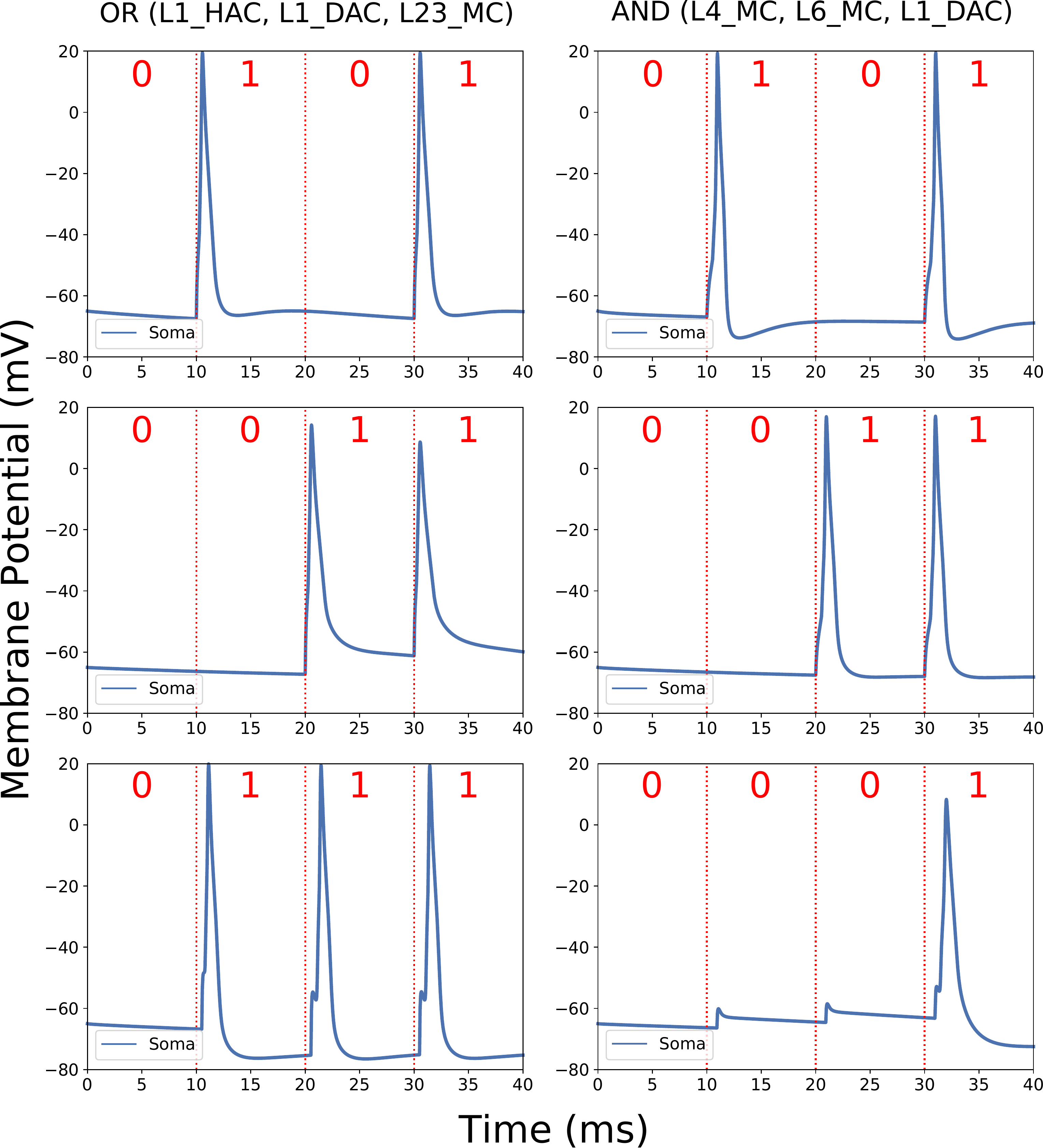}
    \caption{Basic simulation with inputs $[0, 0]$, $[0, 1]$, $[1, 0]$ and $[1, 1]$ for both OR and AND gates with a 10-ms time slot. Inputs 1 and 2 are the first and second rows respectively, the last row is the output.}
\label{fig:ook_mod}
\end{figure}

In this work, a simple \emph{On-Off Keying (OOK)} modulation is implemented where a spike is considered as a bit `$1$' and its absence a bit `$0$' in each time slot (usually 5 ms long) for the inputs into the synthetic gates. The example spikes that propagate along each neuron of a gate is illustrated in Fig.~\ref{fig:ook_mod}. 
When reproducing a $[1, 1]$ input with both L1-HAC and L1-DAC cells (Fig.~\ref{fig:ook_mod} left side), the spikes should arrive at L23-MC with minimum amount of time shift between the spikes to avoid  misprocessing of the inputs by the output cell. 


\subsection{Queuing Theory in Neuronal Circuits}
\label{subsec:queue_theory}

Queueing theory is applied in our analysis with the objective of evaluting the response time and accuracy of the proposed neuronal logic gates. When looking into the times of arrival of spikes, in other words, considering only the electrical behaviour of an electro-chemical synapse, even though there are two inputs, we assume that there is only one queue at the server in which the inputs arrive at a unified rate. At any given moment, only one impulse is carried by the cell and any impulse coming at a rate higher than the service rate may be lost, otherwise, the cell may be able to carry the stimulus and fire again if the input is strong enough to trigger an action potential. The server utilization over a certain period of time, however, depends on the rate of the impulse arrival to the presynaptic terminal.

\subsubsection{Queueing Analysis}
\label{subsubsec:synaptic_queue}

Consider three neuronal cells arranged as a gate, as illustrated in Fig.~\ref{fig:gate3d}, in which two of them are inputs 1 and 2, respectively, and the third cell is the output. We assume that inputs 1 and 2 have \textit{poissonic} rates of $\lambda_{1}$ and $\lambda_{2}$ spikes per second, respectively, and the output cell ``processes'' those inputs with a rate of $\mu$ spikes per second.

Let's also consider that each input has an individual inter-spike interval, $\Delta I_{1}$ and $\Delta I_{2}$, and an inter-neuronal spike interval between both inputs, $\Delta I_{N}$. It is safe to assume that from the perspective of the output cell, the inputs have a merged rate, $\lambda$, defined as~\cite{cowan_1979}

\begin{equation}
\label{eq:unified_rate}
    \lambda = \lambda_{1} + \lambda_{2},
\end{equation}

\noindent
which means that there is only one input with rate $\lambda$ and, analogously, $\Delta I_{N}$ as a unified inter-spike interval. In other words, inputs arrive at time $t^{(f)} + k \cdot \Delta I_{N}$, as depicted in (\ref{eq:syn_th}), where $k$ is a zero-indexed order of arrival.

\begin{figure}[ht]
    \centering
    \includegraphics[width=.4\columnwidth]{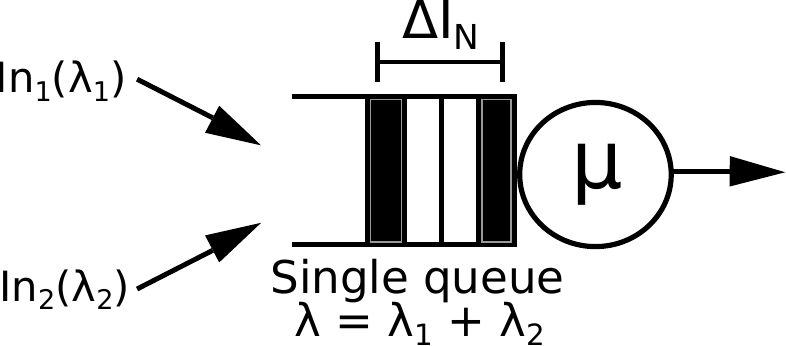}
    \caption{Illustration of merged rates of pre-synpatic spikes into a single queue to be processed by a single server as described in Section~\ref{subsubsec:synaptic_queue}. $\Delta I_{N}$ may have different values for OR and AND gates.}
    \label{fig:queue}
\end{figure}

The system now looks like a single-queue and single-server (Fig.~\ref{fig:queue}). However, if $\mu < \lambda$, there will be no waiting time and, any spike that is not processed at a first come first serve basis, will be lost.

In the case of an OR gate, the output cell should fire when either of the inputs or even both of them fire, hence

\begin{equation}
    \Delta I_{1, 2} \geq 2 \cdot \Delta I_{N},
\end{equation}

\noindent
on the other hand, a cell working as an AND gate should fire only when both inputs fire together. So

\begin{equation}
    \Delta I_{1, 2} \leq Rp + \Delta I_{N}.
\end{equation}

\noindent
where $Rp$ is the refractory period of the output cell. There are different rules for the value of $\Delta I_{N}$ when either input fires (OR gate) or both inputs fire (both gates), thus

\begin{equation}
    \label{eq:or_queue}
    \begin{dcases}
        \Delta I_{N} \geq Rp, & \text{either input fires,}\\
        0 \leq \Delta I_{N} \leq t_{s}, & \text{both inputs fire,}
    \end{dcases}
\end{equation}

\noindent
where $t_{s}$ is some threshold in milliseconds allowing the cell to process neighbouring spikes as $[1, 1]$ input.

If an input arrives with time $t^{(f)}$ (\ref{eq:gsyn}), then the probability of another input arriving \textbf{before} $t^{(f)} + t_{s}$ is
\begin{equation}
\label{eq:arrival_and}
    P(1 | [t^{(f)}, t^{(f)} + t_{s}]) = 1 - e^{-\lambda \Delta I_{N}},
\end{equation}

\noindent
where for an AND gate, the smaller the $t_{s}$, the better to evoke a spike in the output. In the case of an OR gate, we are also interested in another input arriving \textbf{after} $t^{(f)} + t_{s}$, which will transform into

\begin{equation}
\label{eq:arrival_or}
    P(1 | [t^{(f)}, t^{(f)} + t_{s}]) = e^{-\lambda \Delta I_{N}}.
\end{equation}

Using both (\ref{eq:arrival_and}) and (\ref{eq:arrival_or}), it is possible to predict the output and then calculate the accuracy in relation to the expected output. This accuracy should be compared to the approach for calculating the difference between the actual output of the gate and the expected output. The model is further validated in Subsection~\ref{subsec:lgperformance}, where we employ the Mean Squared Error (MSE) analysis.

\section{Simulation Model}
\label{sec:simulation}

In this section, the simulation model for a single neuronal logic gate as well as the application case-study scenario for the suppression of epilepsy, are presented.

\subsection{Single Gate}

The single neuronal logic gates were simulated in two ways. First, they were individually analyzed and simulated in isolation, where their respective accuracy values were evaluated and those results were fitted to the model described in Section~\ref{subsubsec:synaptic_queue}. In isolated form, their configuration is illustrated in Fig.~\ref{fig:gate3d}.

For all simulations, intrinsic parameters of the cell were kept at their default values (such as the length and diameter of each compartment of the cell), and all other parameters of the simulator required to reproduce the desired behaviour are shown in Table~\ref{tab:sim_param}.

\begin{table}[ht]
\centering
    \caption{Parameters for simulation.}
    \label{tab:sim_param}
    \begin{tabular}{c|c}
        \hline
        \hline
        \multicolumn{2}{c}{Simulation Parameters}\\
        \hline
        Synaptic weight & 0.04 $\mu$S \\
        Simulation time & 1 s \\
        Time slot & 5 ms \\
        NetStim.noise & 1 \\
        ExpSyn.tau & 2 \\
        NetCon.threshold (AND) & -64 mV \\
        NetCon.threshold (OR) & 5 mV \\
        NetCon.delay & 0 \\
        Threshold (spike detection) & 0 mV\\ 
        \hline
        $Rp$ & 5 ms\\
        $t_{s}$ & 0 ms\\
        \hline
        \hline
    \end{tabular}
\end{table}

The values of $Rp$ and $t_{s}$ are only used when simulating the queue model and they are not part of the simulation of the network.

\subsubsection{Accuracy}
\label{subsubsec:Accuracy_of_the_gates}

All gates were tested in terms of accuracy with variations in a few parameters to test their performance. These parameters include their firing rate, $\lambda$ and synaptic threshold, $th$. As mentioned earlier in Subsection~\ref{subsec:single_logic_gate_combination}, we are using an OOK modulation to discretize the spiking activity into binary code. Action potentials can shift and get slightly delayed during propagation, and this is due to axonal characteristics. This emphasizes the importance of having a time slot with a fair length of time so there is a fair distinction between different input combinations. The accuracy will measure how correct is the bit train from the output cell with regards to the ideal output that would be generated by an error-free logic gate. In the simulations, random spike trains following a \textit{Poisson} process were stimulated. For each simulation, since the input is random, the number of spikes fired between both inputs with the same rate is approximately the same.

The accuracy is calculated according to the following equation~\cite{hanisch2017digital}:

\begin{equation}
\label{eq:acc}
A(E[Y];Y) = \frac{P_{1,1}+P_{0,0}}{P_{1,1}+P_{1,0}+P_{0,1}+P_{0,0}}
\end{equation}

\noindent
where $P_{Y,E[Y]}$ is the probability of $Y$ given $E[Y]$ in which $Y$ is the actual output and $E[Y]$ is the expected one and $Y\,\&\,E[Y] \in \left \{{0,1}\right \}$.

\subsubsection{Mean Squared Error (MSE)}
\label{subsubsec:mse}

To validate the model proposed in Section~\ref{subsubsec:synaptic_queue}, we estimated how far away our predictions were from the values of accuracy by using \emph{Mean Squared Error}. MSE is a way to measure the quality of an estimator, and in our case, we want to determine the effectiveness of our model with respect to the real accuracy of the gate operation.

Consider that $A$ is the actual accuracy obtained from the real output and that $\overline{A}$ is the predicted accuracy estimated by our model, then the MSE for each point can be calculated as

\begin{equation}
    MSE = \frac{1}{a} \sum_{i=1}^{a} (A_{i} - \overline{A}_{i})^{2},
\end{equation}

\noindent
where $a$ is the number of accuracy values.

\subsection{Neuronal Activity Behavior During Epileptic Seizures}
Neuronal synchronization is the basis for fundamental brain processes. In neurological diseases, such as epilepsy, neuronal synchronicity as well as the balance between excitation and inhibition in populations of cortical neurons can be modified~\cite{alvaradorojas2013, Du2019}. 

In this work, a study on brain seizures is conducted by simulating the activity of neurons grouped in a cortical column and by reproducing the stages of spikes before, during, after, and recovery periods of epileptic seizures with a dynamic firing rate. The frequencies for stimulation varies according to results published by Alvarado-Rojas \textit{et al}~\cite{alvaradorojas2013}. In a simulation with time $T = 1000$ ms, the firing rate $\gamma$ (spikes/s) of the network under epilepsy for the periods of before, during, after, and resting period is based on the following values

\begin{equation}
    \gamma = 
    \begin{dcases}
        30, & \text{if } T \leq 300,\\
        70, & \text{if } 300 < T \leq 650,\\
        180, & \text{if } 650 < T \leq 750,\\
        10, & \text{if } T > 750,
    \end{dcases}
\end{equation}

\noindent
where the evoked activity of the network also follows a \textit{Poisson} process. This procedure can also be described with the pseudo-code presented in Algorithm~\ref{alg:seizure}.

\begin{algorithm}
    \caption{Development of Epileptic Seizure Model}
    \label{alg:seizure}
    \begin{algorithmic}[1]
        \Inputs{$\gamma = \{30, 70, 180, 10\}$\\$T = \{0, 300, 650, 750, 1000\}$}
        \Initialize{$\mathcal{C} = \{c_{1}, ..., c_{n}\}$}
        \For {$c_{n} \in \mathcal{C}$}
            \For {i : 0 $\rightarrow$ length($\gamma$)}
                \State stimulate $c_{n}$ at $(\gamma_{i}, [T_{i}, T_{i+1}])$
            \EndFor
        \EndFor
    \end{algorithmic}
\end{algorithm}

\begin{figure}[t]
    \centering
    \includegraphics[width=.5\columnwidth]{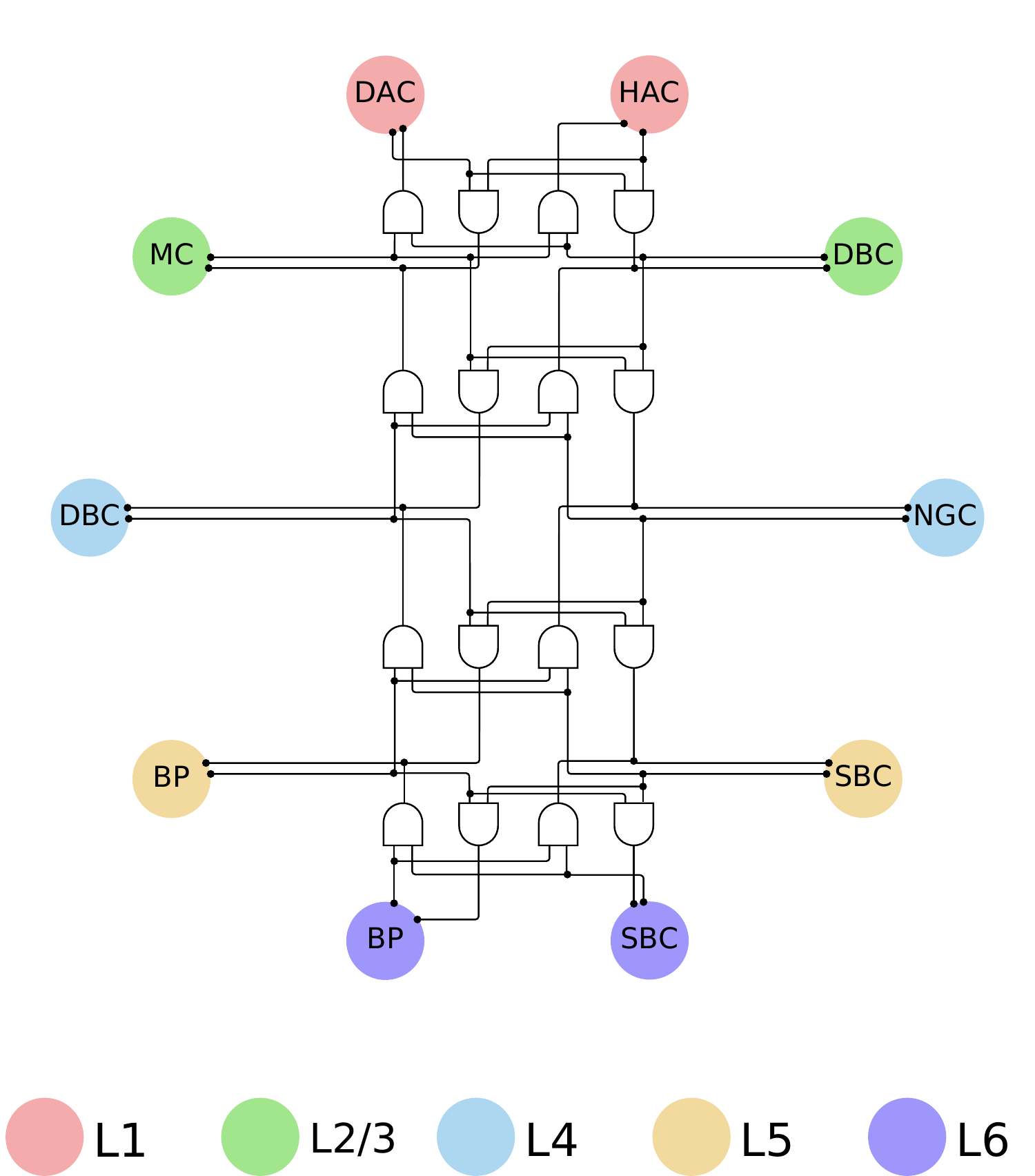}
    \caption{Schematic of the connection of 16 neuronal logic gates in between the 10 cells that forms the network. The placement of a logic gate in the network require the breakage of the natural connections between the cells.}
    \label{fig:gates_network}
\end{figure}

Neuronal AND logic gates were inserted inside a network with 10 neurons (two neurons per cortical layer) that simulated different stages of an epileptic seizure. The positioning of the gates inside the network is depicted in Fig.~\ref{fig:gates_network}.

\section{Results and Discussions}

In this section, we present a discussion over the results for the logic gate analysis and its epilepsy case study.

\subsection{Logic Gate Performance}
\label{subsec:lgperformance}
\begin{figure}[t]
	\centering
	\subfigure[\label{fig:and_acc}]{\includegraphics[width=0.5\columnwidth]{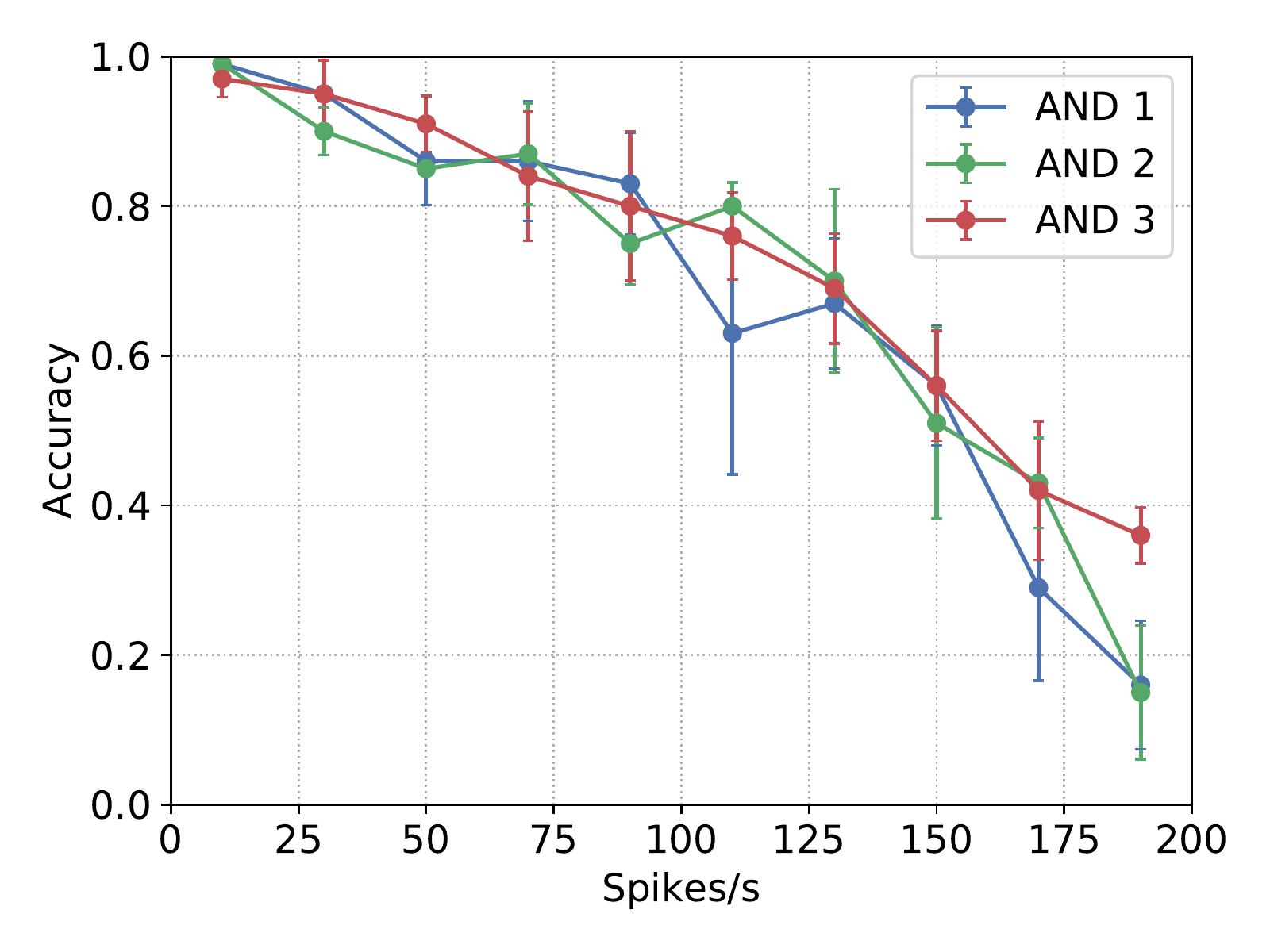}}
	\\
   \subfigure[\label{fig:or_acc}]{\includegraphics[width=0.5\columnwidth]{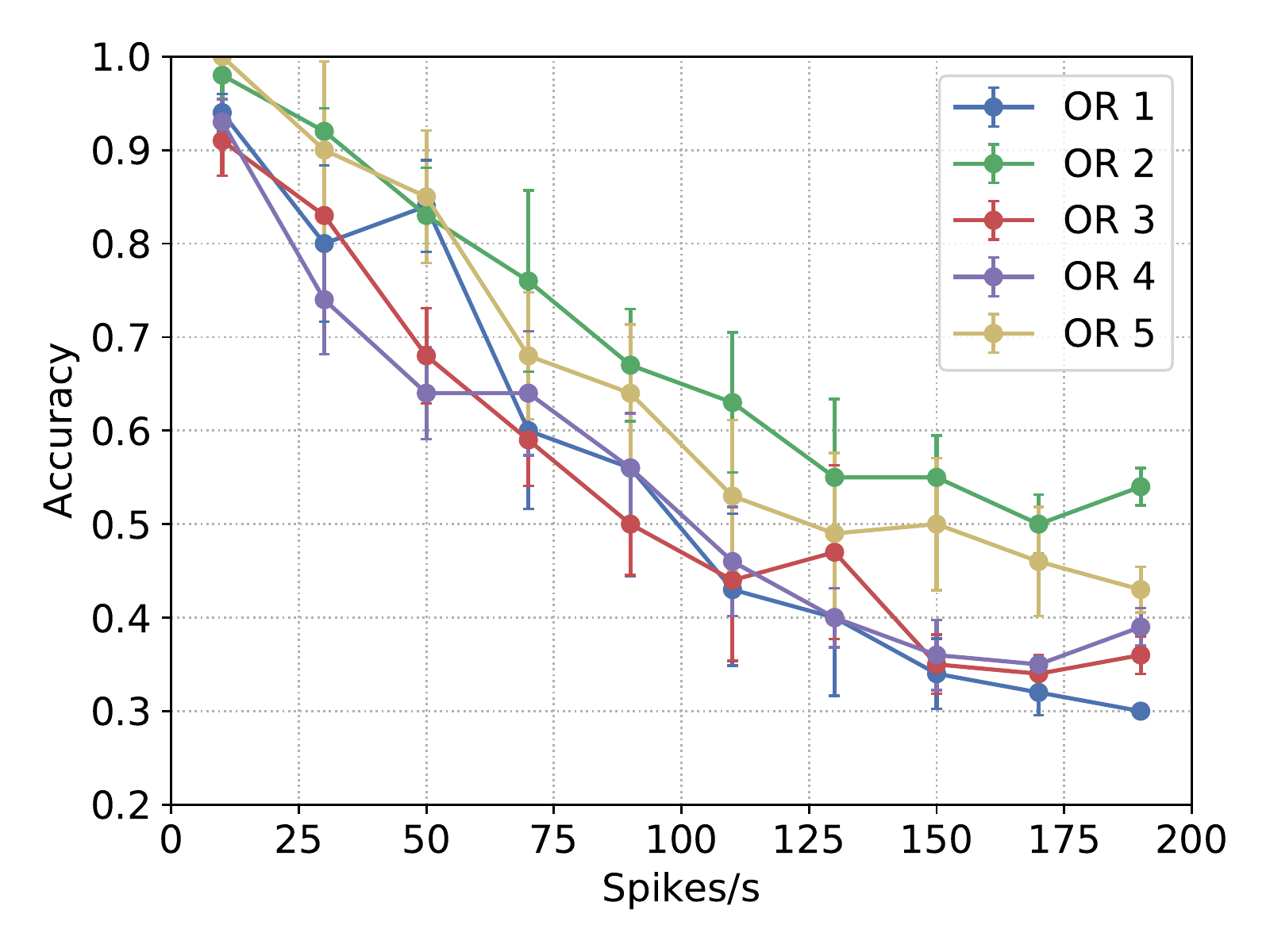}}
    \caption{Mean and standard deviation of the accuracy for the (a) three AND gates and (b) five OR gates. Five simulations were performed for each rate and the firing of the spikes follows a \textit{Poisson} process.\label{fig:rate_acc}}
\end{figure}

For the single gate performance accuracy, we use the configurations that are presented in Fig. 4(a). In the simulations with isolated gates, two different analysis were performed. First, the spiking rate was increased and the accuracy of the gates was computed, and the simulations used the parameters shown in Table~\ref{tab:sim_param}. In both Figs.~\ref{fig:and_acc} and \ref{fig:or_acc}, the accuracy decrease as the firing rate increases, but they are decreasing at different rates due to their different behaviours as shown in Table~\ref{tab:truth_table} and Fig.~\ref{fig:ook_mod}. Even though the different versions of both types of gates have very similar values of accuracy, in Fig.~\ref{fig:and_acc} all of the AND gate configurations have very similar behaviour, this may be due to the fact that, as depicted in Table~\ref{tab:truth_table}, there is only one way for the gate to fire which decreases the chances of misprocessing the inputs. Meanwhile, since OR gates have more ways of firing an output (Table~\ref{tab:truth_table}), the different arrangements may be affecting the processing of the inputs by the output considering that the gates have a bit less similar perfomance between each other when compared to the performance of AND gates. Fig.~\ref{fig:or_acc} shows that OR 2 slightly stands out in performance with better accuracy compared to the other OR gates. To obtain the mean and standard deviation, the simulation ran five times.

\begin{figure}[t]
	\centering
	\subfigure[\label{fig:3dand}]{\includegraphics[width=0.5\columnwidth]{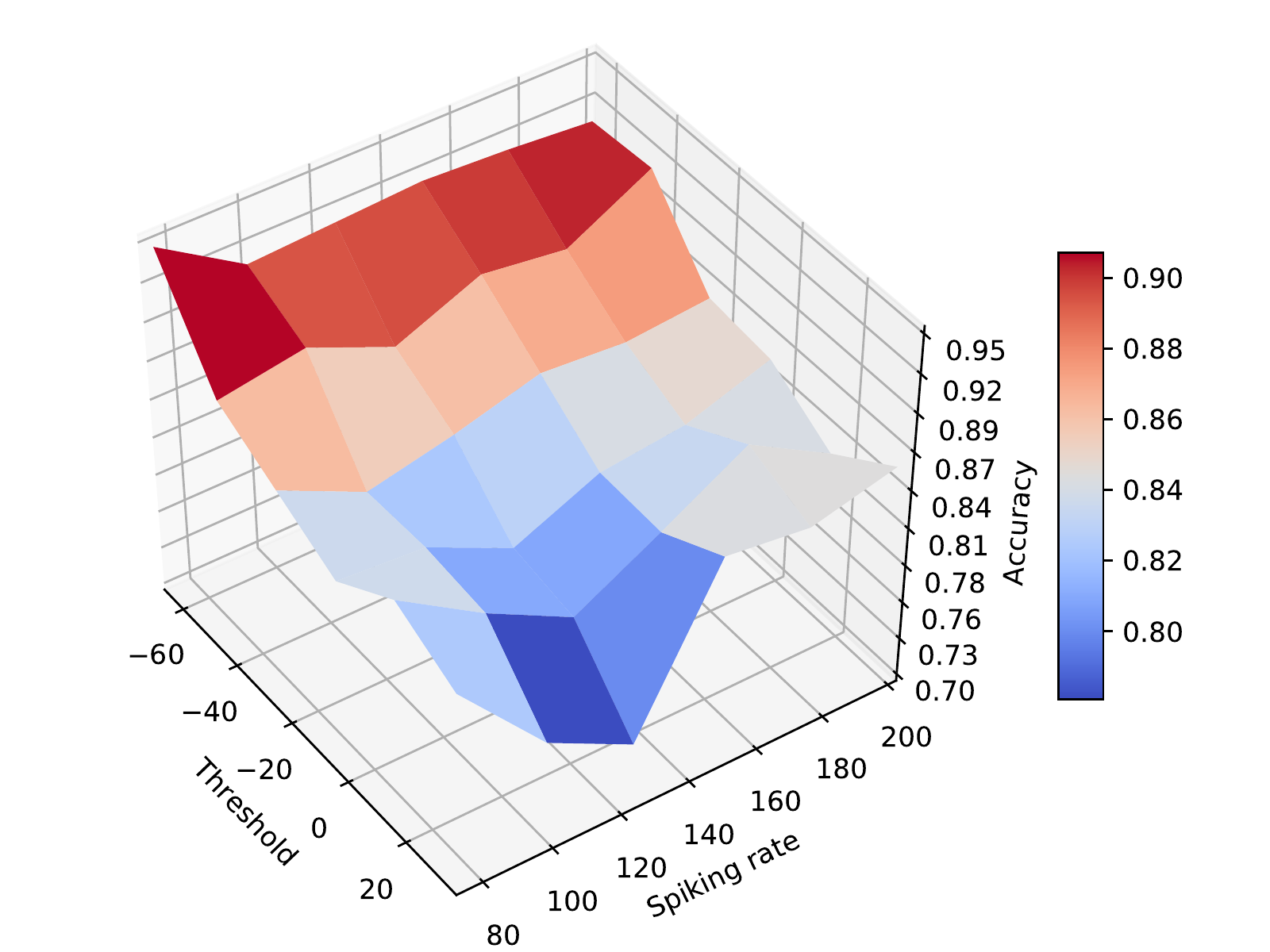}}
	\\
   \subfigure[\label{fig:3dor}]{\includegraphics[width=0.5\columnwidth]{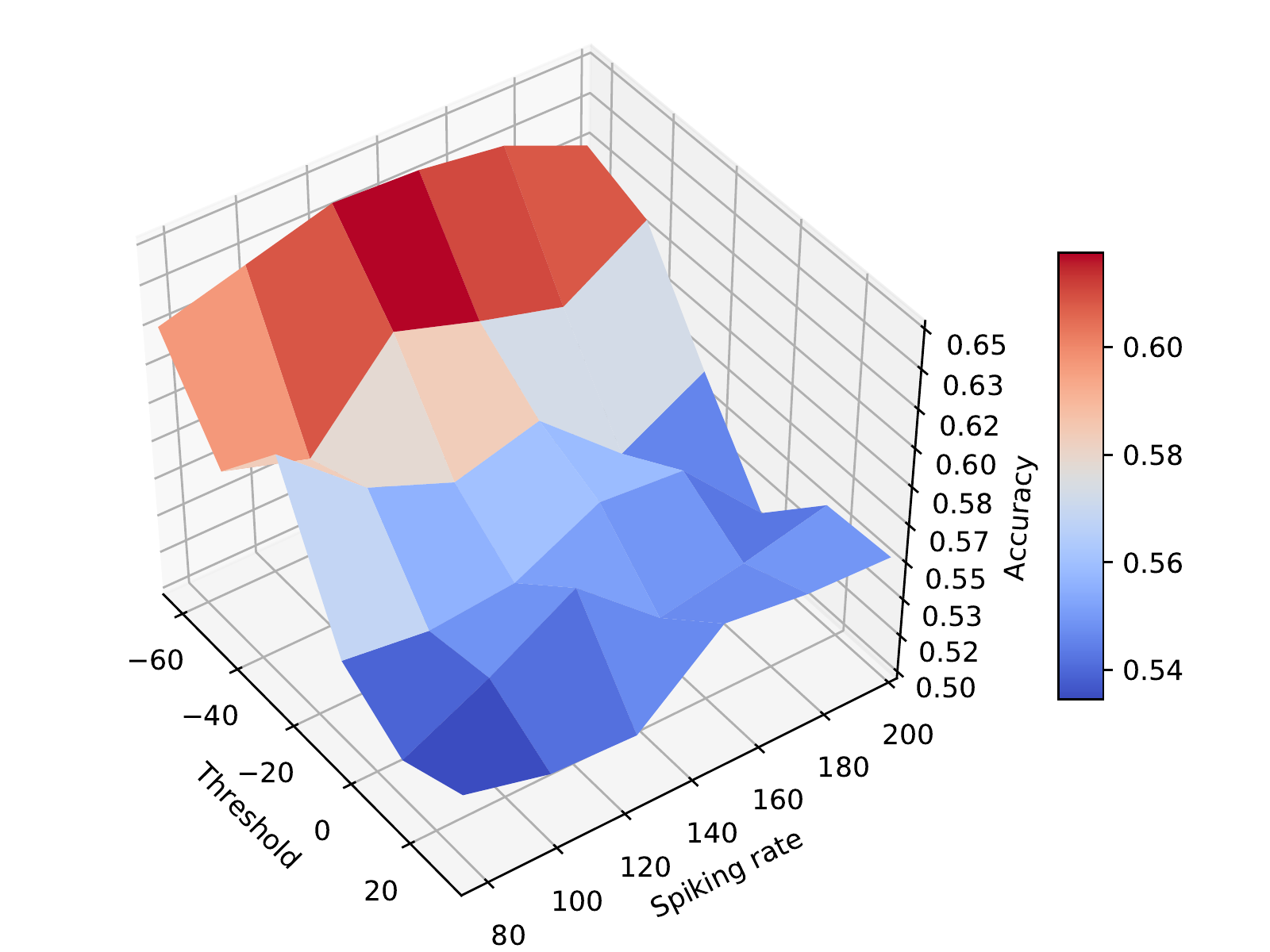}}
    \caption{Spiking rate $\lambda$, threshold $th$ and mean accuracy $Acc$ for gates (a) AND1 and (b) OR2. \label{fig:3d_plots}}
\end{figure}

Second, two gates were picked out, one of each type (OR2 because it showed a better accuracy in Fig.~\ref{fig:or_acc} and, AND1 was actually picked at random since all AND gates have a similar performance), and the accuracy was computed based on both the accuracy and the threshold at the pre-synaptic compartment. In Fig.~\ref{fig:3d_plots}, it is noticeable that the accuracy has a relationship with the threshold, $th_{pre}$ in (\ref{eq:syn_th}), and spiking rate of the neurons, $\lambda_{1, 2}$ in (\ref{eq:unified_rate}).  Generally speaking, lower spiking rates with higher thresholds negatively affects the accuracy of both types of gates. The results presented is based on the mean from three simulation runs. 



The effect of shifting the synchronization of the spikes by up to 4 $ms$ was also analyzed. The results, depicted in Table~\ref{tab:delay_acc}, did not show any specific trend when the shift was increased. For the AND gates, the accuracy remained above 95\%, which represent a difference of at least  9\% in relation to OR gates where the highest performance is approximately 86\%. The highest standard deviation for an AND gate is still over 50 times smaller than the highest value for an OR gate, as aforementioned, this may be due to their different gating behaviours.

\begin{table*}[t]
\centering
    \caption{Accuracy mean and standard deviation values for different delays between inputs.}
    \label{tab:delay_acc}
    \begin{tabular}{c|c|c|c|c|c|c|c|c}
        \hline
        \hline
        \multicolumn{9}{c}{Effect of Delay on Accuracy}\\
        \hline
        \textbf{Types} & AND 1 & AND 2 & AND 3 & OR 1 & OR 2 & OR 3 & OR 4 & OR 5 \\
        \hline
        \textbf{Mean} & $0.9660$ & $0.9530$ & $0.9954$ & $0.7595$ & $0.8595$ & $0.7630$ & $0.7615$ & $0.7720$ \\
        \hline
        \textbf{Std Dev} & $0.00577$ & $0.00604$ & $0.00603$ & $0.02871$ & $0.02752$ & $0.03059$ & $0.03405$ & $0.03039$ \\
        \hline
        \hline
    \end{tabular}
\end{table*}

\begin{figure}[t]
	\centering
	\subfigure[\label{fig:model_and1}]{\includegraphics[width=0.3\columnwidth]{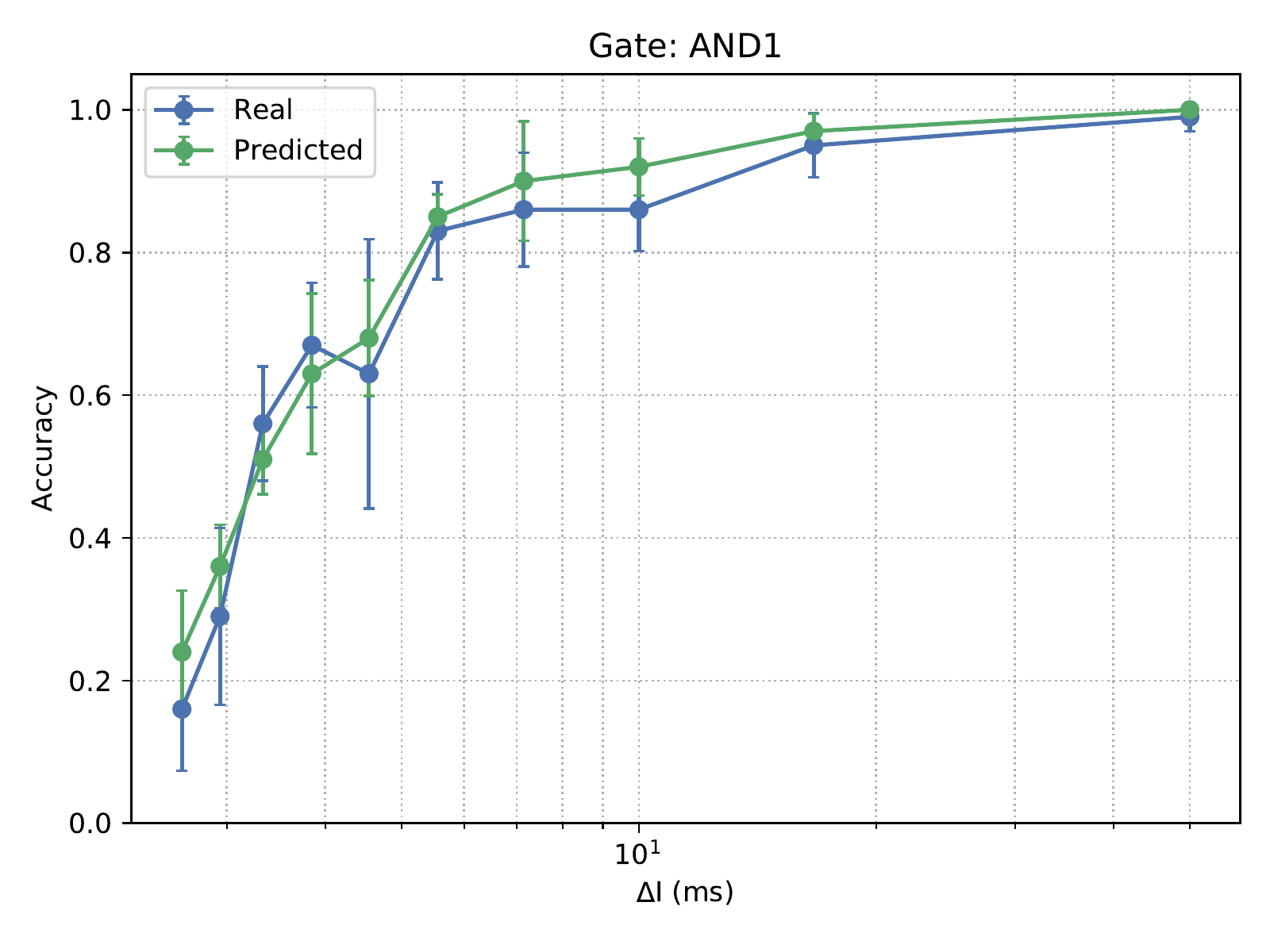}}
	~
   \subfigure[\label{fig:model_OR2}]{\includegraphics[width=0.3\columnwidth]{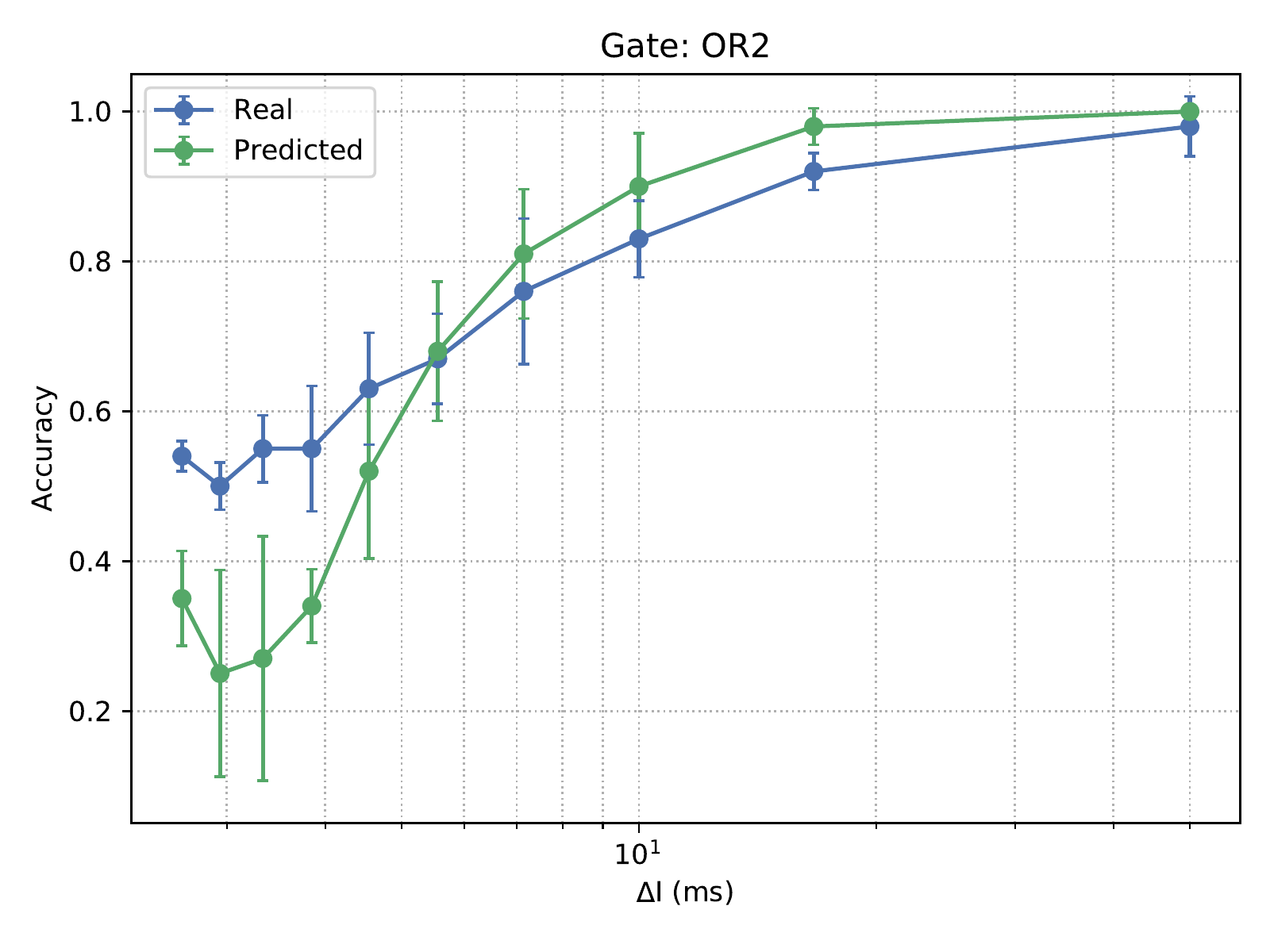}}
    \caption{Real and predicted mean and standard deviation of the accuracy for an (a) AND gate type 1 and (b) OR gate type 2 in relation to the ISI. Predicted results obtained from the implementation of the model proposed in Section~\ref{subsubsec:synaptic_queue}. \label{fig:model_mse}}
\end{figure}

\begin{figure*}[t]
	\centering
	\subfigure[1 ms.\label{fig:mseand1ms}]{\includegraphics[width=0.3\columnwidth]{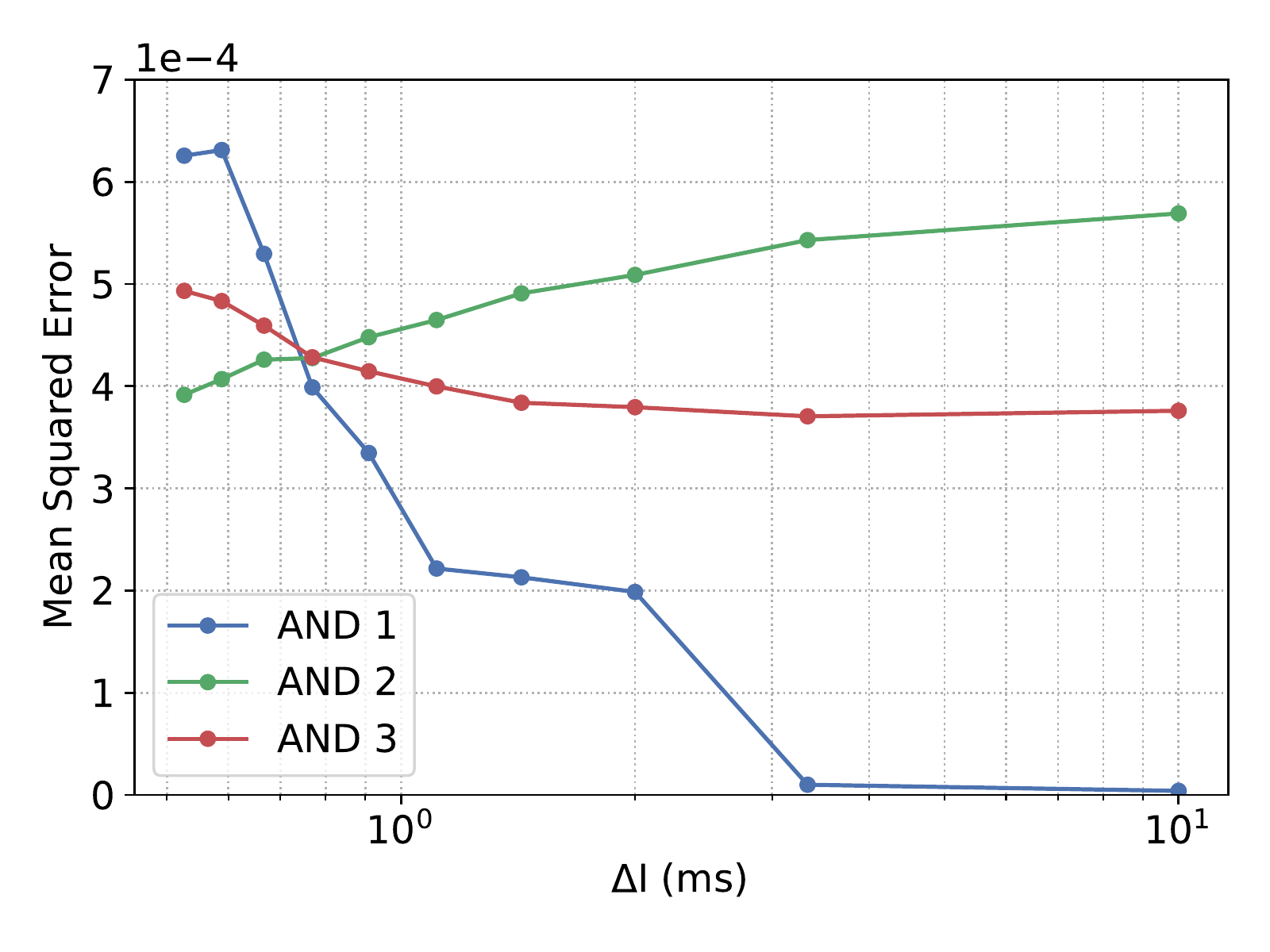}}
	~
	\subfigure[3 ms.\label{fig:mseand3ms}]{\includegraphics[width=0.3\columnwidth]{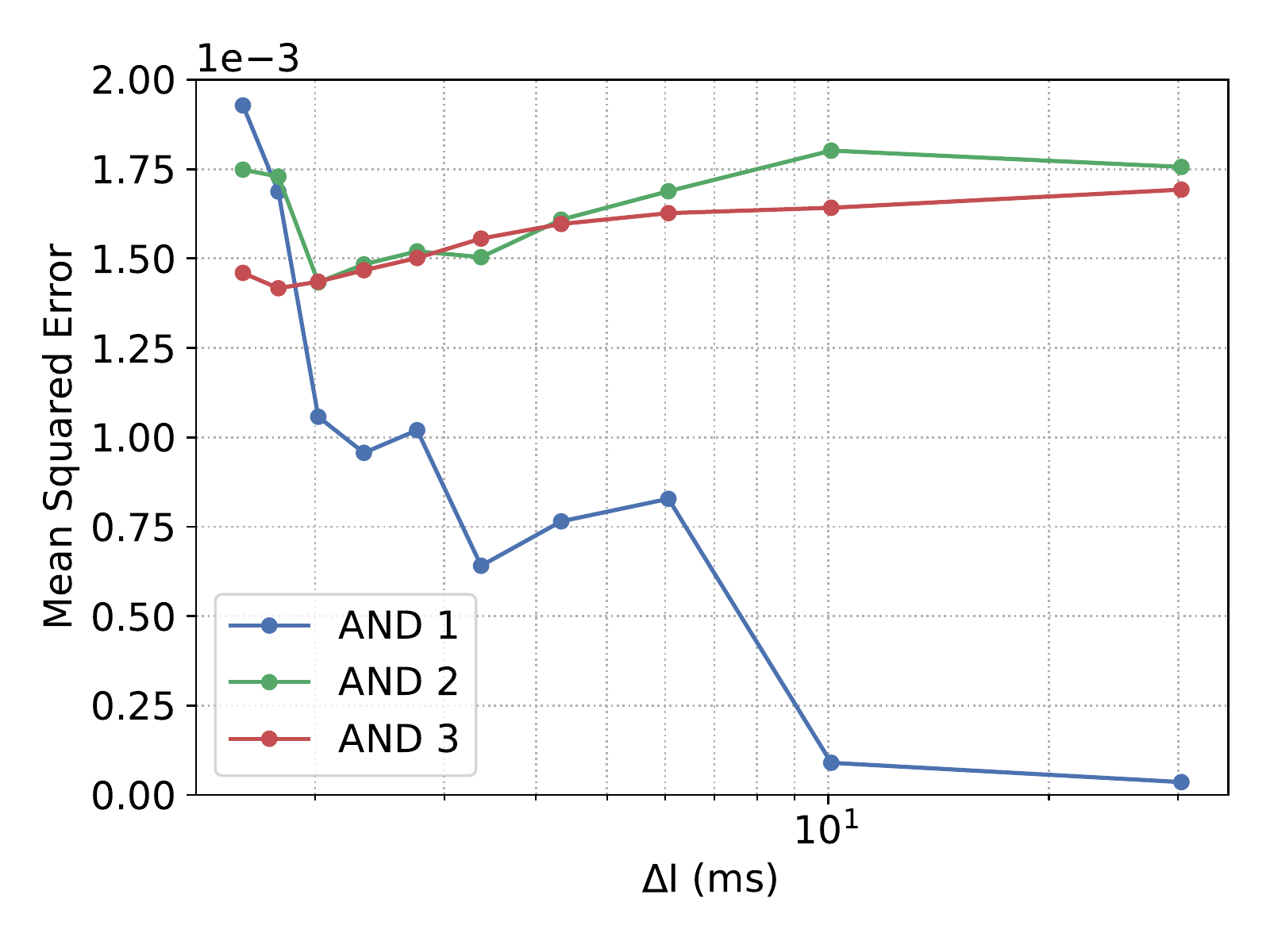}}
	~
	\subfigure[5 ms.\label{fig:mseand5ms}]{\includegraphics[width=0.3\columnwidth]{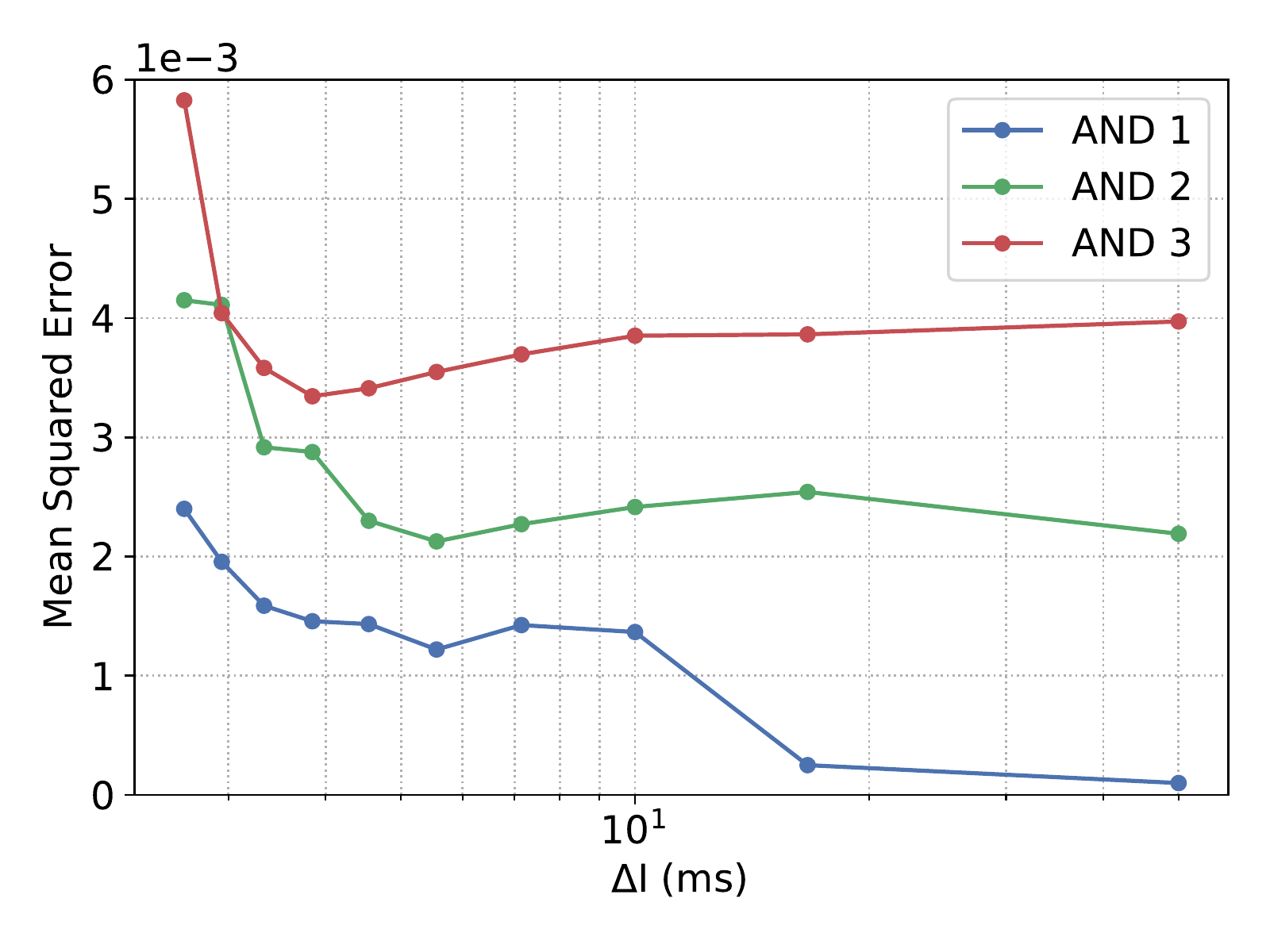}}
	\\
	\subfigure[1 ms.\label{fig:mseor1ms}]{\includegraphics[width=0.3\columnwidth]{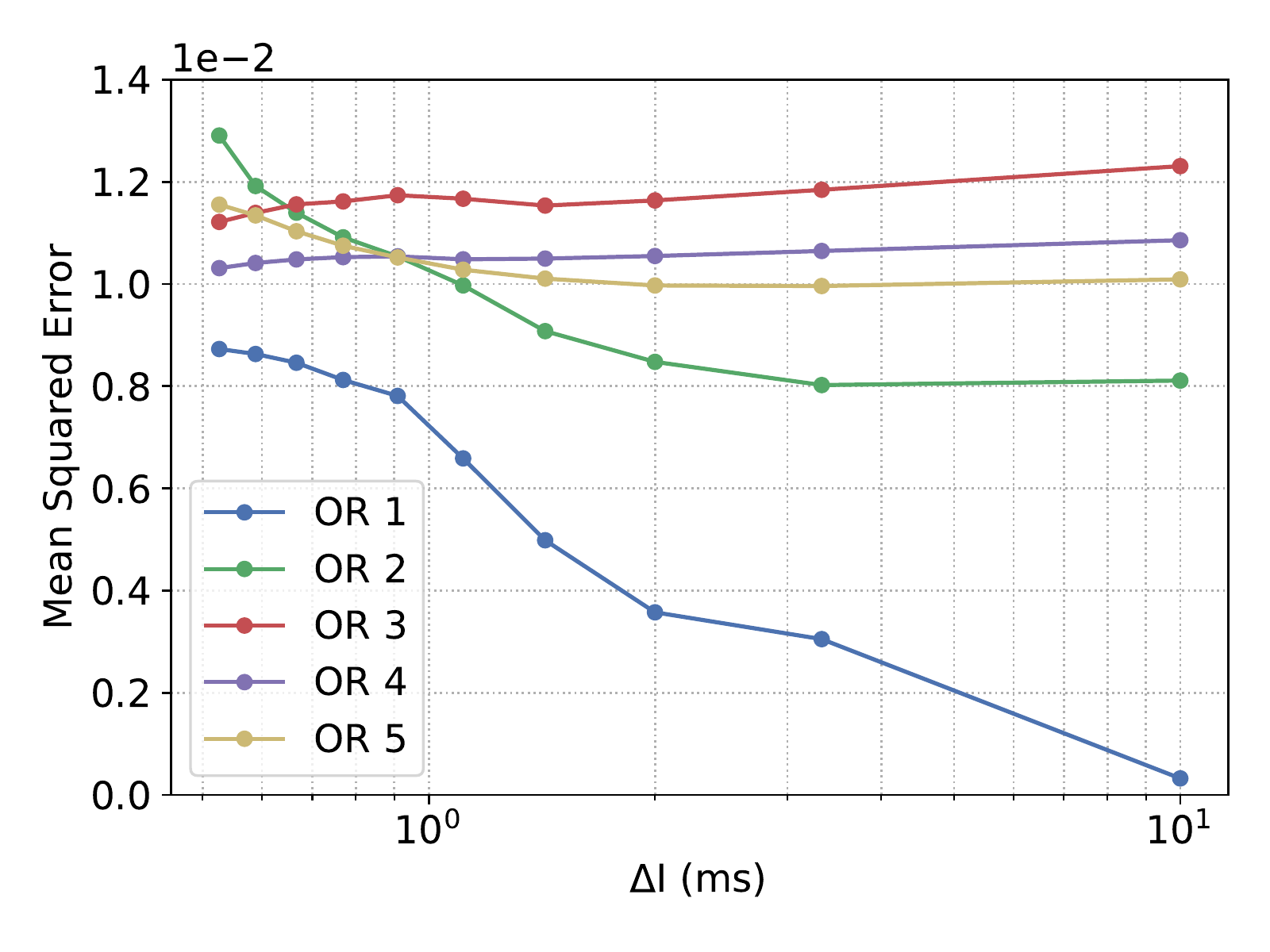}}
	~
    \subfigure[3 ms\label{fig:mseor3ms}]{\includegraphics[width=0.3\columnwidth]{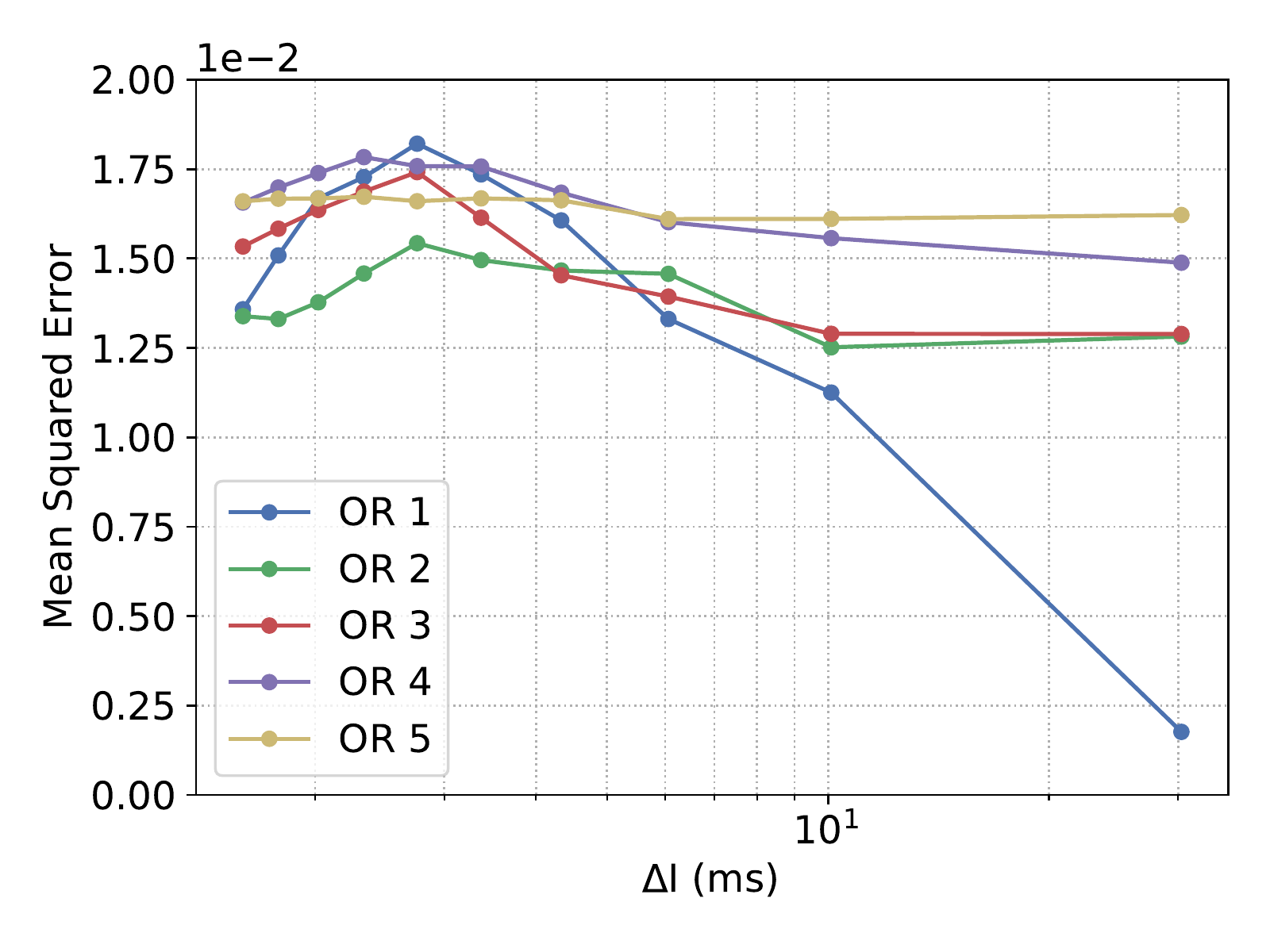}}
    ~
   \subfigure[5 ms.\label{fig:mseor5ms}]{\includegraphics[width=0.3\columnwidth]{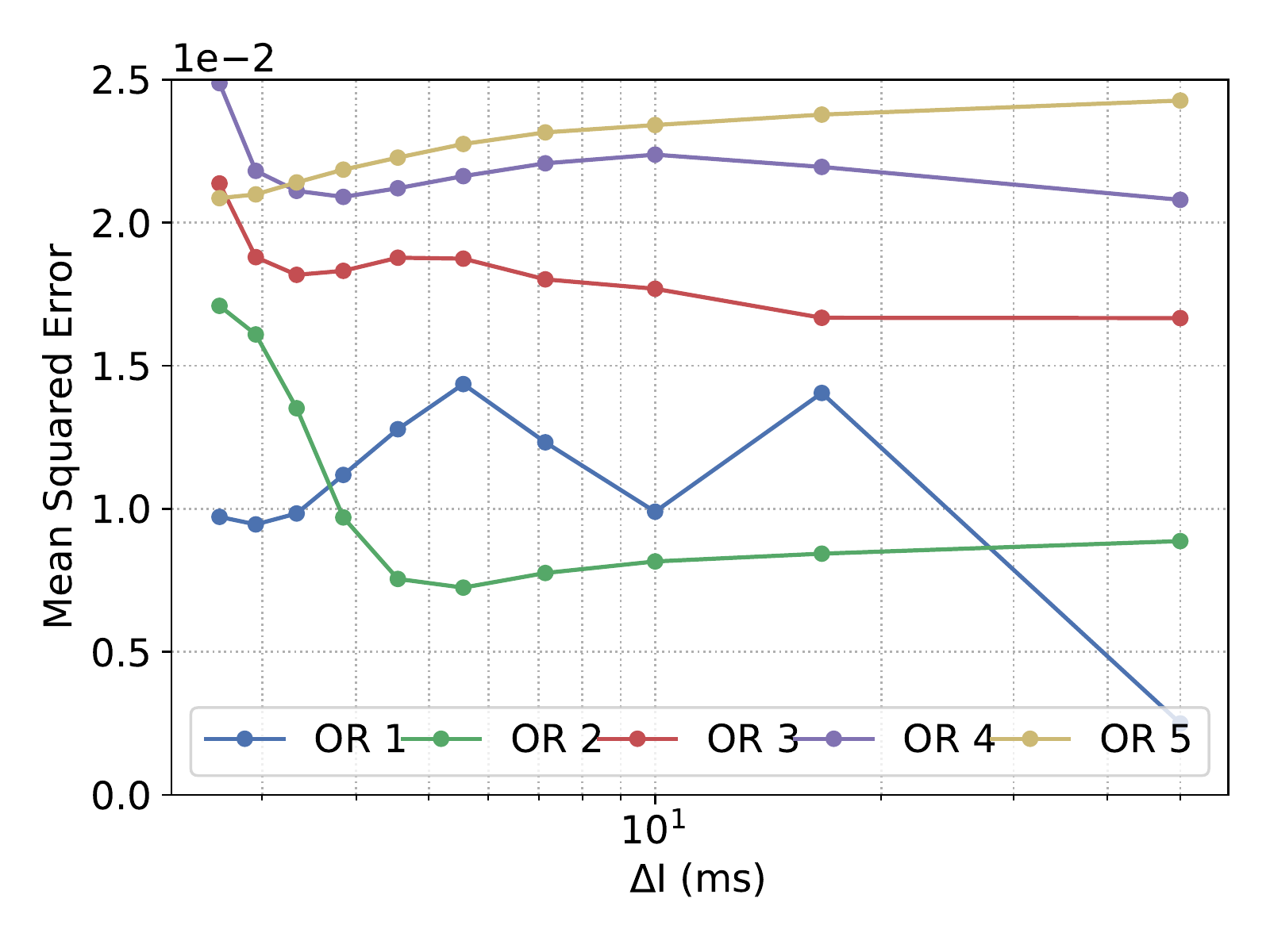}}
    \caption{Mean Squared Error between the predicted accuracy by the model described in Section~\ref{subsubsec:synaptic_queue} and accuracy calculated with real output. Spike trains were sampled at different time slots. Figures~\ref{fig:mseand1ms}, \ref{fig:mseand3ms} and \ref{fig:mseand5ms} show the MSE for all AND gates and Figures~\ref{fig:mseor1ms}, \ref{fig:mseor3ms} and \ref{fig:mseor5ms} show the MSE for all OR gates.\label{fig:mse}}
\end{figure*}

The prediction model was analyzed in relation to $\Delta I_{N}$ and its performance is presented for two gates \emph{AND1} (Fig.~\ref{fig:model_and1}) and \emph{OR2} (Fig.~\ref{fig:model_OR2}). These accuracy values were calculated with a five-millisecond time slot for the discretization analysis of the output cells. The other performances were omitted to avoid redundancy but MSE values are presented in Fig.~\ref{fig:mse} for all of the eight built gates. The results presented in Fig. ~\ref{fig:model_mse} shows that for the AND gates, the model results in slighlty higher accuracy compared to the OR gate. This difference between types of gates reveals itself for the other six gates, four of the OR type and two of the AND type.

In Fig.~\ref{fig:mse}, the sampling frequency was changed to evaluate how the shift among bits from both inputs affect the results of our model in comparison with the real firing of the gates.

Figs.~\ref{fig:mseand1ms}-\ref{fig:mseand5ms} shows the MSE for AND gates with $\Delta I_{N}$ equal to 1, 3, and 5 ms respectively. On the other hand, Figs.~\ref{fig:mseor1ms}-\ref{fig:mseor5ms} show results for the same analysis but OR gates. In both scenarios, the best  MSE was for a time slot with a length of 1 ms, even though an action potential takes a longer period than the firing as well as duration to pass the absolute refractory period. For all gate models, the MSE is quite low showing the robustness of developing gates from the various types of neurons. 

\begin{figure*}[t]
  \centering
  \subfigure[Zero gates in the network.\label{fig:raster_no_gates}]{\includegraphics[width=0.455\columnwidth]{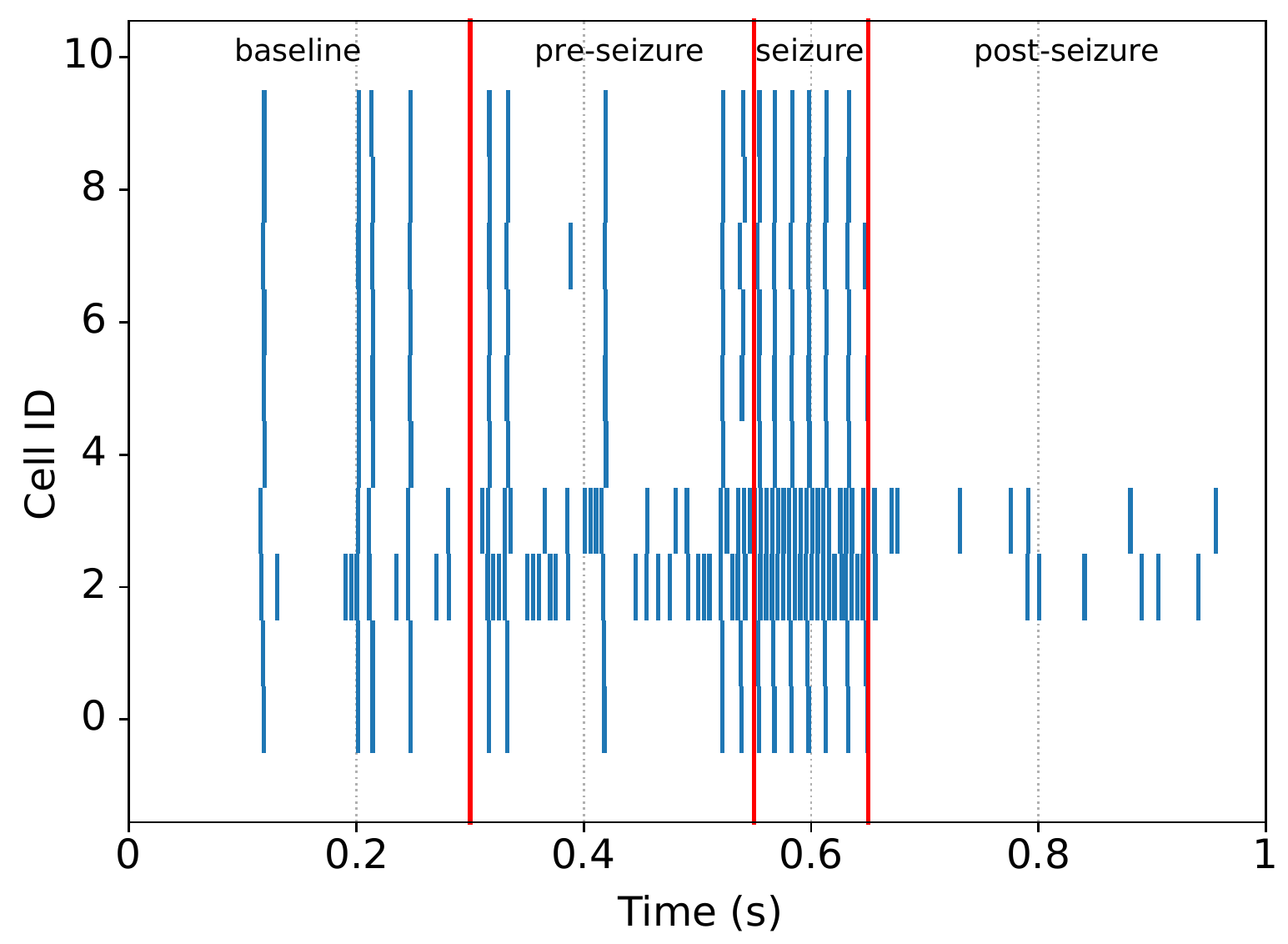}}
  ~
   \subfigure[16 gates in the network.\label{fig:raster_16_gates}]{\includegraphics[width=0.455\columnwidth]{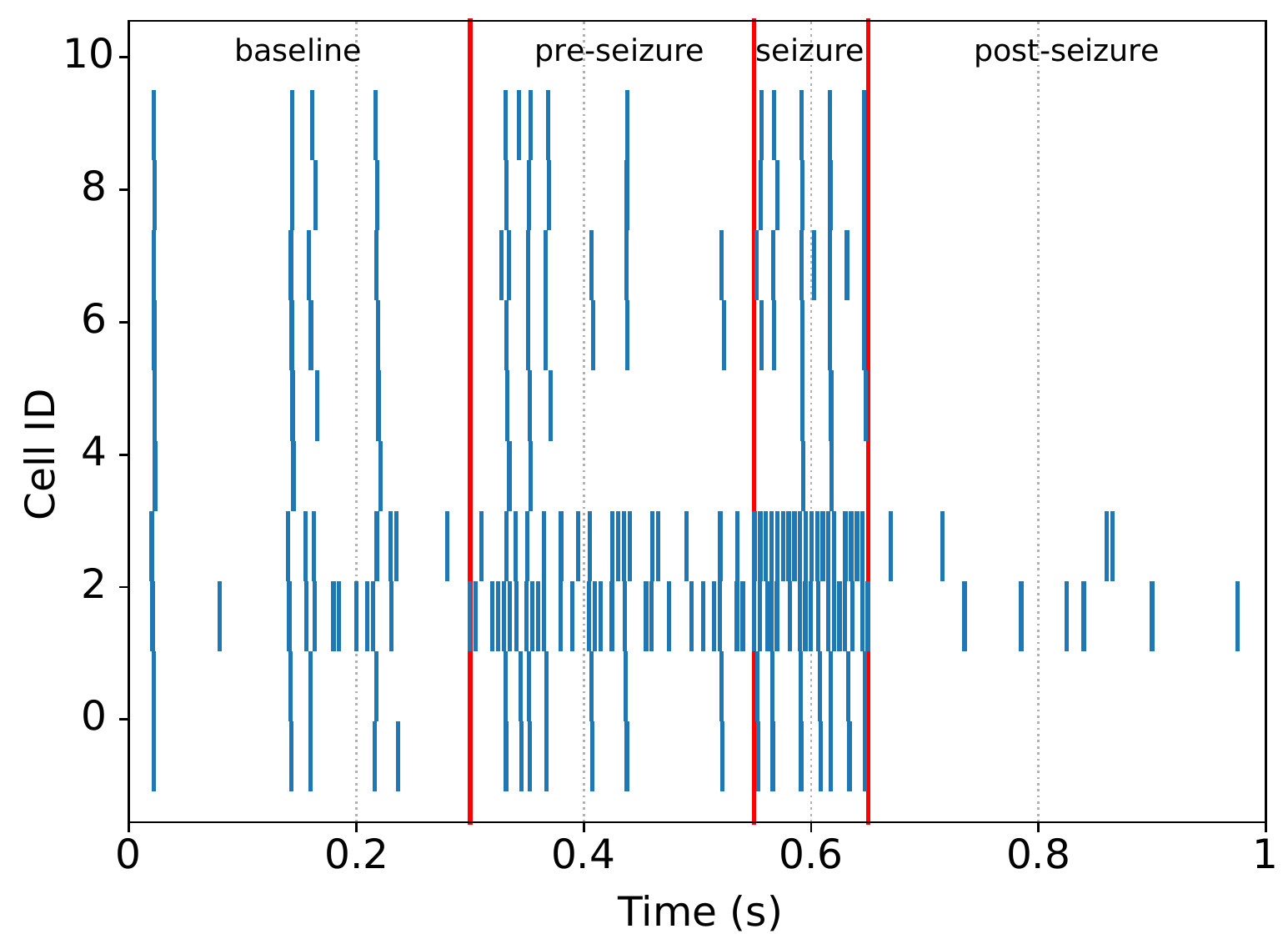}}
   ~
   \subfigure[Mean firing rate of the network.\label{fig:mean_rate}]{\includegraphics[width=0.455\columnwidth]{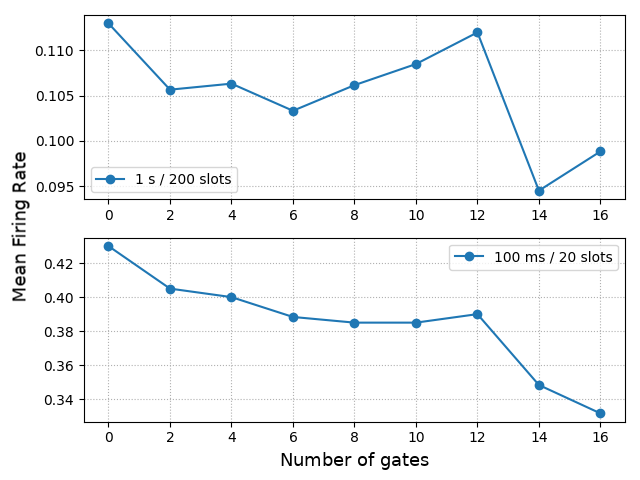}}
    \caption{Simulation of epileptic seizures in a network with 10 neurons (2 neurons per cortical layer), stimulation performed in cells at layer 2/3. (a) Raster plot of the network with no gates inserted and natural connections only as illustrated in Fig.~\ref{fig:connectivity}; (b) Raster plot of the network with 16 neuronal logic gates, natural connections are broken where gates are placed. Placement depicted in Fig.~\ref{fig:gates_network}; (c) Mean firing rate in the network as the more and more gates are placed within it; top graph shows the firing rate of the whole network for all stages as shown in Figures~\ref{fig:raster_no_gates} and \ref{fig:raster_16_gates}; bottom graph shows the firing rate for the whole network but only for the seizure stage.\label{fig:epilepsy_gates}}
\end{figure*}

\subsection{Epilepsy Case Study}
When simulating a network of cells susceptible to epileptic seizures, the analysis was performed in both cases (with and without logic gates within the network). Only one type of AND gate was used in random positions within the network (as illustrated in Fig.~\ref{fig:gates_network}) that is composed of L23-MC, L23-NBC and L1-HAC cells. The placement of the AND gates was chosen at random and we start placing the gates inside the network at regions with high connections to other cells. 

Figure~\ref{fig:mean_rate} shows how placing higher number of gates inside the network may help filter out high frequencies of firing by decreasing the average firing rate of the network. This effect is visually shown in Figures~\ref{fig:raster_no_gates} and \ref{fig:raster_16_gates}, where during the seizure, the entire network resulted in lower levels of activity with 16 gates in comparison with the network that did not contain any gates.



\section{Conclusion}
\label{sec:conclusions}
Even though around 50 million people worldwide have epilepsy, it is estimated that 10\% of the world population will have a seizure during their lifetime without even being an epileptic person. In this paper, the performance of neuronal logic gates was presented as isolated units and their positive effect by smoothing out the synchronous activity of several brain cells that occur during seizures. This approach requires that the cells involved in the gating process should be synthetically engineered and strategically positioned depending on the network connectivity with the objective of improving results.

Also in this work, it was proposed a model based on queue theory concepts that can predict how accurate a specific gating unit can be based on the input and threshold of the output cell. The model showed, in the worst scenario, for OR gates, an MSE of 0.025 while for AND gates this value is of 0.006. The results also show that the sampling frequency of the spike train plays a role in the accuracy of the gates and the quality of the model.
Although this paper only concentrated in the treatment of seizures in the brain, logic gates can also be applied for the encoding of information and have the potential to use synthetic biology to create medical nano-machines to improve the quality of life of people with neurodegenerative diseases and the enhancement of information processing inside the brain.


\bibliographystyle{IEEEtran}
\bibliography{references}

\begin{thebibliography}{10}
\providecommand{\url}[1]{#1}
\csname url@samestyle\endcsname
\providecommand{\newblock}{\relax}
\providecommand{\bibinfo}[2]{#2}
\providecommand{\BIBentrySTDinterwordspacing}{\spaceskip=0pt\relax}
\providecommand{\BIBentryALTinterwordstretchfactor}{4}
\providecommand{\BIBentryALTinterwordspacing}{\spaceskip=\fontdimen2\font plus
\BIBentryALTinterwordstretchfactor\fontdimen3\font minus
  \fontdimen4\font\relax}
\providecommand{\BIBforeignlanguage}[2]{{%
\expandafter\ifx\csname l@#1\endcsname\relax
\typeout{** WARNING: IEEEtran.bst: No hyphenation pattern has been}%
\typeout{** loaded for the language `#1'. Using the pattern for}%
\typeout{** the default language instead.}%
\else
\language=\csname l@#1\endcsname
\fi
#2}}
\providecommand{\BIBdecl}{\relax}
\BIBdecl

\bibitem{nakano2017}
T.~{Nakano}, ``Molecular communication: A 10 year retrospective,'' \emph{IEEE
  Transactions on Molecular, Biological and Multi-Scale Communications},
  vol.~3, no.~2, pp. 71--78, June 2017.

\bibitem{akyildiz2015internet}
I.~F. Akyildiz, M.~Pierobon, S.~Balasubramaniam, and Y.~Koucheryavy, ``The
  internet of bio-nano things,'' \emph{IEEE Communications Magazine}, vol.~53,
  no.~3, pp. 32--40, 2015.

\bibitem{Felicetti2014}
L.~{Felicetti}, M.~{Femminella}, G.~{Reali}, P.~{Gresele}, M.~{Malvestiti}, and
  J.~N. {Daigle}, ``{Modeling CD40-Based Molecular Communications in Blood
  Vessels},'' \emph{IEEE Transactions on NanoBioscience}, vol.~13, no.~3, pp.
  230--243, Sep. 2014.

\bibitem{FELICETTI201627}
\BIBentryALTinterwordspacing
L.~Felicetti, M.~Femminella, G.~Reali, and P.~Liò, ``{Applications of
  molecular communications to medicine: A survey},'' \emph{Nano Communication
  Networks}, vol.~7, pp. 27 -- 45, 2016. [Online]. Available:
  \url{http://www.sciencedirect.com/science/article/pii/S1878778915000411}
\BIBentrySTDinterwordspacing

\bibitem{Balevi2013}
E.~Balevi and O.~B. Akan, ``{A Physical Channel Model for Nanoscale Neuro-Spike
  Communications},'' \emph{IEEE Transactions on Communications}, vol.~61,
  no.~3, pp. 1178--1187, March 2013.

\bibitem{Thorpe2001}
S.~Thorpe, A.~Delorme, and R.~V. Rullen, ``{Spike-based strategies for rapid
  processing},'' \emph{Neural Networks}, vol.~14, no.~6, pp. 715 -- 725, 2001.

\bibitem{Rolls2011}
E.~T. Rolls and A.~Treves, ``{The neuronal encoding of information in the
  brain},'' \emph{Progress in Neurobiology}, vol.~95, no.~3, pp. 448 -- 490,
  2011.

\bibitem{Rinkus2010}
G.~Rinkus, ``{A cortical sparse distributed coding model linking mini- and
  macrocolumn-scale functionality},'' \emph{Frontiers in Neuroanatomy}, vol.~4,
  p.~17, 2010.

\bibitem{Rao2011}
Y.~Huang and R.~P.~N. Rao, ``{Predictive coding},'' \emph{Wiley
  Interdisciplinary Reviews: Cognitive Science}, vol.~2, no.~5, pp. 580--593,
  2011.

\bibitem{Luczak2015}
A.~Luczak, B.~L. McNaughton, and K.~D. Harris, ``{Packet-based communication in
  the cortex},'' \emph{Nature Reviews Neuroscience -- Perspectives}, vol.~16,
  pp. 1 -- 11, Oct 2015.

\bibitem{Zeldenrust2018}
F.~Zeldenrust, W.~J. Wadman, and B.~Englitz, ``{Neural Coding With Bursts --
  Current State and Future Perspectives},'' \emph{Frontiers in Computational
  Neuroscience}, vol.~12, p.~48, 2018.

\bibitem{Adonias2018}
G.~L. Adonias, M.~T. Barros, L.~Doyle, and S.~Balasubramaniam, ``{Utilising EEG
  Signals for Modulating Neural Molecular Communications},'' in \emph{5th ACM
  International Conference on Nanoscale Computing and Communication 2018 (ACM
  NanoCom'18)}, Reykjavik, Iceland, Sep. 2018.

\bibitem{Billimoria5910}
C.~P. Billimoria, R.~A. DiCaprio, J.~T. Birmingham, L.~F. Abbott, and
  E.~Marder, ``Neuromodulation of spike-timing precision in sensory neurons,''
  \emph{Journal of Neuroscience}, vol.~26, no.~22, pp. 5910--5919, 2006.

\bibitem{Yi2014}
G.-S. Yi, J.~Wang, X.-L. Wei, K.-M. Tsang, W.-L. Chan, and B.~Deng, ``Neuronal
  spike initiation modulated by extracellular electric fields,'' \emph{PLOS
  ONE}, vol.~9, no.~5, pp. 1--10, 05 2014.

\bibitem{Choe2016}
J.~Choe, B.~A. Coffman, D.~T. Bergstedt, M.~D. Ziegler, and M.~E. Phillips,
  ``{Transcranial Direct Current Stimulation Modulates Neuronal Activity and
  Learning in Pilot Training},'' \emph{Frontiers in Human Neuroscience},
  vol.~10, p.~34, 2016.

\bibitem{pavon2014}
C.~Agustín-Pavón and M.~Isalan, ``Synthetic biology and therapeutic
  strategies for the degenerating brain,'' \emph{BioEssays}, vol.~36, no.~10,
  pp. 979--990, 2014.

\bibitem{Loscri2015}
V.~{Loscrí} and A.~M. {Vegni}, ``{An Acoustic Communication Technique of
  Nanorobot Swarms for Nanomedicine Applications},'' \emph{IEEE Transactions on
  NanoBioscience}, vol.~14, no.~6, pp. 598--607, Sep. 2015.

\bibitem{Loscri2018}
V.~{Loscrí}, B.~D. {Unluturk}, and A.~M. {Vegni}, ``{A Molecular Optical
  Channel Model Based on Phonon-Assisted Energy Transfer Phenomenon},''
  \emph{IEEE Transactions on Communications}, vol.~66, no.~12, pp. 6247--6259,
  Dec 2018.

\bibitem{Egan2019}
M.~{Egan}, V.~{Loscri}, T.~Q. {Duong}, and M.~{Di Renzo}, ``{Strategies for
  Coexistence in Molecular Communication},'' \emph{IEEE Transactions on
  NanoBioscience}, vol.~18, no.~1, pp. 51--60, Jan 2019.

\bibitem{akyildiz2008nanonetworks}
I.~F. Akyildiz, F.~Brunetti, and C.~Bl{\'a}zquez, ``Nanonetworks: A new
  communication paradigm,'' \emph{Computer Networks}, vol.~52, no.~12, pp.
  2260--2279, 2008.

\bibitem{Adonias2019}
G.~L. Adonias, A.~Yastrebova, M.~T. Barros, S.~Balasubramaniam, and
  Y.~Koucheryavy, ``{A Logic Gate Model based on Neuronal Molecular
  Communication Engineering},'' in \emph{Proceedings of the 4th Workshop on
  Molecular Communications}, Linz, Austria, Apr. 2019.

\bibitem{hasty2002engineered}
J.~Hasty, D.~McMillen, and J.~J. Collins, ``Engineered gene circuits,''
  \emph{Nature}, vol. 420, no. 6912, p. 224, 2002.

\bibitem{Goldental2014}
A.~Goldental, S.~Guberman, R.~Vardi, and I.~Kanter, ``A computational paradigm
  for dynamic logic-gates in neuronal activity,'' \emph{Frontiers in
  Computational Neuroscience}, vol.~8, p.~52, 2014.

\bibitem{Vogels10786}
T.~P. Vogels and L.~F. Abbott, ``Signal propagation and logic gating in
  networks of integrate-and-fire neurons,'' \emph{Journal of Neuroscience},
  vol.~25, no.~46, pp. 10\,786--10\,795, 2005.

\bibitem{morse2002time}
D.~Morse and P.~Sassone-Corsi, ``Time after time: inputs to and outputs from
  the mammalian circadian oscillators,'' \emph{Trends in neurosciences},
  vol.~25, no.~12, pp. 632--637, 2002.

\bibitem{jirsa2014}
V.~K. Jirsa, W.~C. Stacey, P.~P. Quilichini, A.~I. Ivanov, and C.~Bernard,
  ``{On the nature of seizure dynamics},'' \emph{Brain}, vol. 137, no.~8, pp.
  2210--2230, 06 2014.

\bibitem{Markram2015}
H.~Markram \emph{et~al.}, ``{Reconstruction and Simulation of Neocortical
  Microcircuitry},'' \emph{Cell}, vol. 163, no.~2, pp. 456--492, 2015.

\bibitem{McCulloch1943}
W.~S. McCulloch and W.~Pitts, ``A logical calculus of the ideas immanent in
  nervous activity,'' \emph{The bulletin of mathematical biophysics}, vol.~5,
  no.~4, pp. 115--133, Dec 1943.

\bibitem{Miyamoto2013}
\BIBentryALTinterwordspacing
T.~Miyamoto, S.~Razavi, R.~DeRose, and T.~Inoue, ``{Synthesizing
  Biomolecule-Based Boolean Logic Gates},'' \emph{ACS Synthetic Biology},
  vol.~2, no.~2, pp. 72--82, 2013, pMID: 23526588. [Online]. Available:
  \url{https://doi.org/10.1021/sb3001112}
\BIBentrySTDinterwordspacing

\bibitem{Wang2012}
\BIBentryALTinterwordspacing
R.-S. Wang, A.~Saadatpour, and R.~Albert, ``{Boolean modeling in systems
  biology: an overview of methodology and applications},'' \emph{Physical
  Biology}, vol.~9, no.~5, p. 055001, sep 2012. [Online]. Available:
  \url{https://doi.org/10.1088/1478-3975/9/5/055001}
\BIBentrySTDinterwordspacing

\bibitem{SONG2016380}
T.~Song, P.~Zheng, M.~D. Wong, and X.~Wang, ``Design of logic gates using
  spiking neural p systems with homogeneous neurons and astrocytes-like
  control,'' \emph{Information Sciences}, vol. 372, pp. 380 -- 391, 2016.

\bibitem{wirdatmadja2019analysis}
S.~Wirdatmadja, P.~Johari, A.~Desai, Y.~Bae, E.~K. Stachowiak, M.~K.
  Stachowiak, J.~M. Jornet, and S.~Balasubramaniam, ``Analysis of light
  propagation on physiological properties of neurons for nanoscale
  optogenetics,'' \emph{IEEE Transactions on Neural Systems and Rehabilitation
  Engineering}, vol.~27, no.~2, pp. 108--117, 2019.

\bibitem{Eckenstein2286}
F.~Eckenstein and M.~Sofroniew, ``{Identification of central cholinergic
  neurons containing both choline acetyltransferase and acetylcholinesterase
  and of central neurons containing only acetylcholinesterase},'' \emph{Journal
  of Neuroscience}, vol.~3, no.~11, pp. 2286--2291, 1983.

\bibitem{Peters2010}
A.~Peters, ``The morphology of minicolumns,'' in \emph{The neurochemical basis
  of autism}.\hskip 1em plus 0.5em minus 0.4em\relax Springer, 2010, pp.
  45--68.

\bibitem{Mishra2019}
A.~Mishra and S.~K. Majhi, ``{A comprehensive survey of recent developments in
  neuronal communication and computational neuroscience},'' \emph{Journal of
  Industrial Information Integration}, vol.~13, pp. 40 -- 54, 2019.

\bibitem{Zhou2018}
S.~Zhou and Y.~Yu, ``{Synaptic E-I Balance Underlies Efficient Neural
  Coding},'' \emph{Frontiers in Neuroscience}, vol.~12, p.~46, 2018.

\bibitem{Carnevale2009}
N.~T. Carnevale and M.~L. Hines, \emph{{The NEURON Book}}, 1st~ed.\hskip 1em
  plus 0.5em minus 0.4em\relax New York, NY, USA: Cambridge University Press,
  2009.

\bibitem{Hines2009}
M.~Hines, A.~Davison, and E.~Muller, ``{NEURON and Python},'' \emph{Frontiers
  in Neuroinformatics}, vol.~3, p.~1, 2009.

\bibitem{pospischil2008minimal}
M.~Pospischil, M.~Toledo-Rodriguez, C.~Monier, Z.~Piwkowska, T.~Bal,
  Y.~Fr{\'e}gnac, H.~Markram, and A.~Destexhe, ``Minimal hodgkin--huxley type
  models for different classes of cortical and thalamic neurons,''
  \emph{Biological cybernetics}, vol.~99, no. 4-5, pp. 427--441, 2008.

\bibitem{hodgkin1952quantitative}
A.~L. Hodgkin and A.~F. Huxley, ``A quantitative description of membrane
  current and its application to conduction and excitation in nerve,''
  \emph{The Journal of physiology}, vol. 117, no.~4, pp. 500--544, 1952.

\bibitem{Akyildiz2019}
I.~F. {Akyildiz}, M.~{Pierobon}, and S.~{Balasubramaniam}, ``{An Information
  Theoretic Framework to Analyze Molecular Communication Systems Based on
  Statistical Mechanics},'' \emph{Proceedings of the IEEE}, vol. 107, no.~7,
  pp. 1230--1255, July 2019.

\bibitem{platkiewicz2010threshold}
J.~Platkiewicz and R.~Brette, ``{A threshold equation for action potential
  initiation},'' \emph{PLoS computational biology}, vol.~6, no.~7, p. e1000850,
  2010.

\bibitem{cowan_1979}
R.~Cowan, ``The uncontrolled traffic merge,'' \emph{Journal of Applied
  Probability}, vol.~16, no.~2, p. 384–392, 1979.

\bibitem{hanisch2017digital}
N.~Hanisch and M.~Pierobon, ``Digital modulation and achievable information
  rates of thru-body haptic communications,'' in \emph{Disruptive Technologies
  in Sensors and Sensor Systems}, vol. 10206.\hskip 1em plus 0.5em minus
  0.4em\relax International Society for Optics and Photonics, 2017, p. 1020603.

\bibitem{alvaradorojas2013}
C.~Alvarado-Rojas, K.~Lehongre, J.~Bagdasaryan, A.~Bragin, R.~Staba, J.~Engel,
  V.~Navarro, and M.~LE~VAN~QUYEN, ``{Single-unit activities during epileptic
  discharges in the human hippocampal formation},'' \emph{Frontiers in
  Computational Neuroscience}, vol.~7, p. 140, 2013.

\bibitem{Du2019}
J.~Du, V.~Vegh, and D.~C. Reutens, ``Small changes in synaptic gain lead to
  seizure-like activity in neuronal network at criticality,'' \emph{Scientific
  Reports}, vol.~9, no.~1, p. 1097, 2019.

\end{thebibliography}

\end{document}